\documentclass[aps,prb,twocolumn,superscriptaddress]{revtex4-1}
\usepackage[margin=1.in]{geometry}
\usepackage{physics}
\newcommand{\hc}[1]{{#1}^{\dagger}}
\usepackage{amsmath}
\usepackage{amsthm}
\usepackage{amsfonts}
\DeclareMathOperator{\spn}{span}

\usepackage{amssymb}
\usepackage{mathtools}
\usepackage{bbold}
\usepackage{bm}
\usepackage{graphicx}
\usepackage{indentfirst}
\usepackage{nameref}
\usepackage{hyperref}
\hypersetup{
	colorlinks,
	citecolor=black,
	filecolor=black,
	linkcolor=black,
	linktoc=all
	urlcolor=black
}

\begin{document}
\title{Effect of the electron-lattice coupling on the charge and magnetic order in rare-earth nickelates}
\author{Stepan Fomichev}
\email{fomichev@physics.ubc.ca}
\affiliation{Department of Physics and Astronomy, University of
British Columbia, Vancouver B.C. V6T 1Z1, Canada} 
\affiliation{Stewart Blusson Quantum Matter Institute, University of
British Columbia, Vancouver B.C. V6T 1Z4, Canada}
\author{Giniyat Khaliullin}
\affiliation{Max Planck Institute for Solid State Research, D-70569 Stuttgart, Germany}
\author{Mona Berciu}
\affiliation{Department of Physics and Astronomy, University of
British Columbia, Vancouver B.C. V6T 1Z1, Canada} 
\affiliation{Stewart Blusson Quantum Matter Institute, University of
British Columbia, Vancouver B.C. V6T 1Z4, Canada}
\date{\today}

\begin{abstract}
We investigate the impact of electron-lattice coupling on the stability of various magnetic orders in rare-earth nickelates. We use the Hartree-Fock approximation, at zero temperature, to study an effective, two-band model with correlations characterized by a Hubbard $U$ and a Hund's $J$. This is coupled to breathing-mode distortions of the octahedral oxygen cages, described semi-classically, with a Holstein term. We analyze the effect of the various parameters on the resulting phase diagram, in particular on the charge disproportionation and on the magnetic order.  We confirm that the coupling to the lattice cooperates with Hund's coupling and thus encourages charge disproportionation. We also find that it favors the fully disproportionated, 4-site periodic magnetic order of type $\Uparrow 0 \Downarrow 0$. Other convergent magnetic phases, such as the collinear $\uparrow\uparrow\downarrow\downarrow$ and non-collinear $\uparrow\rightarrow\downarrow\leftarrow$ states, do not couple to the lattice because of their lack of charge disproportionation. Novel phases, \textit{e.g.} with charge disproportionation but no magnetic order, are also found to be stabilized in specific conditions. 
\end{abstract}

\maketitle

\section{Introduction}

The rare-earth nickelates $R$NiO$_3$, $R$ being any rare-earth element
from La to Lu, are a class of materials that have
generated considerable interest because of their complex phase diagram, where
both a metal-insulator transition (MIT) and a magnetic ordering
transition can be tuned via pressure, strain, and/or variations in the
size of the $R$ ion (except for $R$=La, which is a metal at all
temperatures).\cite{Catalan2008, Medarde1997} Understanding the
features of this phase diagram is of significant interest for the
advancement of our basic knowledge of strongly correlated systems, but
also because of their potential for applications, especially in
heterostructures \cite{Hepting2017} and memory storage.
\cite{Giovannetti2009, Spaldin2017, Scott2007}

The perovskite $R$NiO$_3$ consists of Ni ions arranged on a simple
cubic lattice with lattice constant $a$, and connected through ligand
O, such that each Ni ion is inside an octahedral cage of oxygen atoms. \footnote{These
octahedral cages undergo a variety of tilts, twists and other complicated distortions due to the rare-earth ion being too small to accommodate a true perovskite lattice. However, such distortions change little across the temperature range we are interested in: the main change to lattice structure comes from the breathing mode distortion \cite{Bodenthin2011}. As such, for the purposes of our analysis we can focus on the breathing-mode distortion.} Standard valence counting
suggests the starting configuration to be Ni:$3d^7$. Crystal fields
split the $3d$ levels into well-separated $e_g$ and $t_{2g}$
manifolds, suggesting a doubly-degenerate $t_{2g}^6e_g^1$
configuration that should be unstable to Jahn-Teller distortions.
\cite{Rodriguez-Carvajal1998} Such distortions are not observed
experimentally,\cite{Scagnoli2006} so the $e_g$ degeneracy must be
resolved in some other way. Many scenarios have been proposed,
including, most notably, (i) charge disproportionation (CD),
\cite{Catalan2008, Medarde2009} and (ii) a negative charge transfer
(NCT) energy.\cite{Mizokawa2000, Park2012, Lau2013, Puggioni2012,
  Caviglia2012}

The CD scenario posits that below the MIT there are two
inequivalent Ni sites with different charge, $3d^73d^7\rightarrow
3d^{7-\delta}3d^{7+\delta}$; this is then thought to drive the experimentally
observed distortion \cite{Garcia-Munoz1992a} of the O octahedra into
small/large ones around the two inequivalent Ni sites. As typical for
strongly correlated insulators, magnetic order also develops at
or below the MIT phase line.

In contrast, the NCT scenario has all the Ni in the $3d^8$ ($S=1$)
configuration, with each releasing a ligand hole into the O band. The
resulting 1/6 filled (with holes) O band is metallic at high
temperatures. In this view, the MIT is primarily due to
electron-phonon coupling which distorts the O octahedra into small and
large ones ($pp$ and $pd$ hopping are enhanced on the shorter bonds),
resulting in pairs of ligand holes localized on the small octahedra
and locked into a singlet with their central Ni. The spins of the Ni
ions located in the large cages, on the other hand, order magnetically
at or below the MIT temperature.

The electron-phonon coupling is obviously important for the MIT
transition in both scenarios (even if in rather different ways), but
its impact on the magnetic
order is not well understood. Neutron scattering experiments on powders
\cite{Garcia-Munoz1992b, Munoz2009} indicate a magnetic ordering
wavevector $\mathbf{Q}_m= \frac{\pi}{2a}(1, 1, 1)$ ($\frac{\pi}{2a} (1, 0,
1)$ in perovskite notation), which is half of the value associated
with the lattice distortion ordering (and charge modulation, in the CD
scenario) of $\mathbf{Q}_c= \frac{\pi}{a}(1,1,1)$. The orientations and
magnitudes of the local magnetic moments in the 4-site (counting Ni
only) magnetic unit cell are still under debate. Three
leading contenders are the fully-disproportionated antiferromagnetic
state $\Uparrow 0 \Downarrow 0$, \cite{Haule2017} and two partially disproportionated
orders: the collinear \cite{Lee2011} order $\Uparrow\uparrow\Downarrow\downarrow$, and the
non-collinear \cite{Scagnoli2006, Lu2018} order $\Uparrow\rightarrow\Downarrow\leftarrow$.  (Throughout this
work we use fat arrows to indicate larger spin magnitudes).

While the preferred magnetic order is likely to be primarily decided
by the electron-electron interactions, as is generally the case in
strongly-correlated systems, it is possible that the strong coupling
to the lattice also plays an important role by favoring or hindering
some of these possible candidates. We study this possibility here
using an effective two-orbital Hamiltonian that, within the
appropriate framework (discussed below), can be used to model both the
CD and the NCT scenarios. Our work builds on that of Lee \textit{et
  al.} \cite{Lee2011} who studied magnetic orders possible in similar
multi-orbital models. The main novelty is that our model also includes
coupling to the lattice, at the semi-classical level. 

We verify that the electron-lattice coupling favors insulating charge
order, as noticed by earlier investigators. While a number of 4-site
magnetic orders, including $\Uparrow 0 \Downarrow 0$, $\uparrow\uparrow\downarrow\downarrow$ and $\uparrow\rightarrow\downarrow\leftarrow$ are found to be
self-consistent within our model, we find that $\Uparrow 0 \Downarrow 0$ is the only
one that has nonzero charge modulation $\delta \neq 0$: as such it is strongly
favored by the electron-lattice coupling. The other
magnetic orders $\uparrow\uparrow\downarrow\downarrow$ and $\uparrow\rightarrow\downarrow\leftarrow$ are only self-consistent when $\delta=0$,
and as a result do not couple to the lattice in our model. 
As the strength of the electron-lattice
interaction increases, we also find that a novel phase is stabilized, wherein charge
modulation occurs in the absence of magnetic order.

The paper is organized as follows: In section II, we describe our
effective model Hamiltonian. Section III reviews the Hartree-Fock
calculation used to study it, as well as its numerical implementation.
In Section IV we present and discuss our results. Finally, Section V
contains  our conclusions.

\section{The Model}\label{sec:model}

We consider a simple cubic lattice, with lattice constant $a$, which we set to 1. At each
site $i$, two $e_g$ ``effective'' orbitals $\ket{z} \equiv \ket{3z^2-r^2}$
and $\ket{\overline{z}} \equiv \ket{x^2-y^2}$ are active, and
$\hc{d}_{iz\sigma}, \hc{d}_{i\bar{z}\sigma}$ are the electronic creation
operators associated with them. The physical connection to the actual material of these
and the other ``effective'' degrees of freedom
that we introduce below is
discussed after the Hamiltonian is fully defined.

The Hamiltonian we study is defined as:
\begin{equation}
\label{Ham}
\hat{H} = \hat{T} + \hat{H}_{e-e} + {H}_{lat} + \hat{H}_{e-lat}.
\end{equation}

The kinetic energy $\hat{T}=\hat{T}_1 + \hat{T}_2 + \hat{T}_4$
includes up to 4$^{\text{th}}$ nearest-neighbor hopping. The hopping term
$\hat{T}_1$ includes hopping between nearest-neighbor $\ket{z}$ orbitals along the
$z$-axis, and between $\ket{x}\equiv \ket{3x^2-r^2}$ and
$\ket{y}\equiv \ket{3y^2-r^2}$ orbitals along the $x$- and $y$-axes,
respectively. There is no nearest-neighbor hopping in the $z$-direction between
$\ket{\overline{z}}$ orbitals, \textit{etc.}, because these orbitals are
orthogonal to their corresponding ligand O. As a result:
\begin{equation}
\hat{T}_1 = -t_1 \sum_{i \sigma} \sum_{\eta=x,y,z}\left(\hc{d}_{i\eta\sigma}
d_{i+\eta,\eta\sigma} + \text{ h.c.} \right).
\end{equation}
This can be easily expressed in terms of the $\hc{d}_{iz\sigma}, \hc{d}_{i\bar{z}\sigma}$ operators using the identities:
  \begin{align}
    &\hc{d}_{ix\sigma} = -\frac{1}{2}\hc{d}_{iz\sigma} + \frac{\sqrt{3}}{2} \hc{d}_{i\bar{z}\sigma}, \nonumber \\
    &\hc{d}_{iy\sigma} = -\frac{1}{2}\hc{d}_{iz\sigma} - \frac{\sqrt{3}}{2} \hc{d}_{i\bar{z}\sigma}.
\label{pol-orb}
  \end{align}
Similarly, we define 2$^{\text{nd}}$ nearest-neighbor and 4$^{\text{th}}$ nearest-neighbor hopping terms
$\hat{T}_2$ and $ \hat{T}_4$, respectively, keeping all such terms that have finite hopping amplitudes. Collecting all terms and after a Fourier transform to the $\mathbf{k}$-space basis, defined as
\begin{equation}
\label{FT}
\hc{d}_{\mathbf{k}a\sigma}= \frac{1}{\sqrt{N}} \sum_{i}^{} e^{i \mathbf{k} \mathbf{R}_i}  \hc{d}_{i a\sigma}
\end{equation}
where $a = z, \bar{z}$, $N\rightarrow \infty$ is the number of sites in the system with periodic boundary conditions, and $\mathbf{k}$ is defined inside the full Brillouin zone $-\pi < k_\eta \le \pi$, $\eta=x,y,z$,  the hopping Hamiltonian is brought to the standard  form
\begin{equation}
\label{TT}
\hat{T} = \sum_{\mathbf{k}ab\sigma} t_{ab}(\mathbf{k})
\hc{d}_{\mathbf{k}a\sigma} d_{\mathbf{k}b\sigma}.
\end{equation}
The coefficients $t_{ab}(\mathbf{k})$ are
listed in the Appendix.

The on-site electron-electron interactions are described by the Kanamori  Hamiltonian: \cite{Kanamori1963, Oles1983, Georgescu2015}
\begin{align}
\nonumber &\hat{H}_{e-e} = U \sum_{ia}
\hat{n}_{ia\uparrow}\hat{n}_{ia\downarrow} + U'\sum_{i\sigma}
\hat{n}_{iz\sigma}\hat{n}_{i\bar{z}\bar{\sigma}} \\ \nonumber &+
(U'-J)\sum_{i\sigma} \hat{n}_{iz\sigma}\hat{n}_{i\bar{z}\sigma} - J
\sum_{i\sigma} d^{\dagger}_{iz\sigma}d_{iz\bar{\sigma}}
d^{\dagger}_{i\bar{z}\bar{\sigma}}d_{i\bar{z}\sigma} \\ &+ J \sum_{ia}
d^{\dagger}_{ia\uparrow}d_{i\bar{a}\uparrow}d^{\dagger}_{ia\downarrow}d_{i\bar{a}\downarrow}.
\end{align}
with the spherically symmetric choice $U' = U - 2J$.
 \cite{Castellani1978} Here, $\hat{n}_{ia\sigma} = d^{\dagger}_{ia\sigma}d_{ia\sigma}$
counts the electrons with spin $\sigma$ in the $a=z,\bar{z}$ orbital at
site $i$, and $ \hat{n}_i = \sum_{a\sigma}^{} \hat{n}_{ia\sigma}$. We use this full
form of $\hat{H}_{e-e}$ as opposed to the simpler one used in Lee
\textit{et al.} \cite{Lee2011} because it leads only to minor
complications in the mean-field treatment and has a formal derivation
based on allowed Coulomb intraionic interactions. \cite{Lu2018, Kanamori1963, Oles1983} This change explains the
quantitative differences between our results -- in the absence of coupling to the lattice -- and those of Lee
\textit{et al.}

Next, ${H}_{lat}$ describes, at the semi-classical level, the
breathing-mode distortion resulting in contracted and expanded
octahedra:
\begin{equation}
\nonumber
{H}_{lat} = \sum_i \left( \frac{k}{2} (\delta U_i)^2 + \frac{A}{4} (\delta U_i)^4
\right)
\end{equation}
where $\delta U_i$ is the (isotropic) change in the Ni-O bond length of the octahedral cage
surrounding site $i$, and we include quartic anharmonicity to ensure
reasonable values for these distortions.

The octahedral distortions affect the on-site electron energies, hence the electron-lattice interaction term:
\begin{equation}
\nonumber
\hat{H}_{e-lat} = -g \sum_i \delta U_i \left( \hat{n}_i - 1 \right).
\end{equation}
It is convenient to use dimensionless variables $u_i = (\delta U_i) k/g$, in
terms of which we rewrite:
\begin{align}
H_{lat} + \hat{H}_{e-lat} = 2\epsilon_b \sum_i \left( \frac{1}{2}
u_i^2 + \frac{\alpha}{4} u_i^4 \right) \nonumber \\ - 2\epsilon_b
\sum_i u_i \left( \hat{n}_i - 1 \right).
\end{align}
where $\epsilon_b = g^2/2k$ is the energy gain from of the breathing-mode distortion for $u_i=1,\alpha=0$, and $\alpha \sim A$ is the dimensionless parameter characterizing the anharmonicity.

To summarize, there are 7 parameters characterizing this Hamiltonian:
the three hoppings $t_1, t_2, t_4$; the on-site Coulomb repulsion $U$
and Hund's exchange $J$; the electron-lattice coupling strength $\epsilon_b$ and the
dimensionless anharmonicity parameter $\alpha$.

Before concluding this section, we comment on how this Hamiltonian
describes the two scenarios discussed above. Within the CD scenario,
only the Ni $e_g$ orbitals are relevant as valence orbitals, so they
should be directly identified with the $e_g$ orbitals of this model.
In this view, the O are electronically inert, and serve only to
modulate the on-site energy at the Ni sites when the cages are
distorted. Even though the octahedra are known to tilt and rotate, the
Ni-O distances stay equal inside each cage, so it is reasonable to use
a single distortion $u_i$ to characterize each octahedron. Note that
while our model does not explicitly impose constraints between
distortions on neighboring cages, the mean-field solution will turn
out to satisfy them, as discussed below.

The relevance to the NCT scenario is less obvious. Here, a
full description of the electronic degrees of freedom include the
Ni $e_g$ orbitals but also the ligand O $2p$ ones, as discussed in
Ref. [\onlinecite{Johnston2014}]. Of course, in principle one could do a mean-field
treatment on the Hamiltonian used there to discuss the MIT, to find
what magnetic order it favors and how (or if) it is affected by the
fact that the O displacements modulate the hopping amplitudes. The
difficulty is that the magnetic unit cell contains 4 Ni together with
their 12 O, \textit{i.e.} 20 distinct orbitals (counting spins as well).
Needless to say, when combined with the multitude of possible
mean-field parameters in such a large basis, the problem becomes
rather unwieldy.

On the other hand, when considering a single Ni plus its O octahedron,
one finds that the relevant eigenstates on the O sites are linear
combinations with the same $e_g$ symmetries like the atomic Ni
orbitals. This is because in order for an electron from the O band to
move in the $e_g$ Ni manifold (and thus leave behind a ligand hole),
it has to come from an O state that will hybridize (via $t_{pd}$
hopping) with the Ni orbital, and that only occurs if they have the
same point symmetry. This is what allows us to identify the two $e_g$
``effective'' orbitals as being these O-based linear combinations with
the correct $e_g$ symmetry, surrounding various Ni sites, and into
which the ligand hole can go.\cite{Subedi2015} The complication here is that such
orbitals centred about nearest-neighbour Ni ions are not orthogonal, so the true
``effective'' orbitals must correct for that and are therefore
somewhat more extended and more complicated than simple linear
combinations of O orbitals from each octahedral cage. As a result, all hoppings
and electronic parameters are now likely strongly
renormalized from their atomic values. We do not attempt to estimate
their realistic values: instead, we will treat them as free parameters.
This allows us to investigate what kind of magnetic orders arise in
different regions of this large parameter space, and thus cover
simultaneously both the CD and the NCT scenarios (their parameters are
likely to be quite different).

In a broader view, the CD and NCT scenarios are limiting cases in a
continuum of possibilities. The $t_{pd}$ hopping always leads to some
hybridization between the Ni $e_g$ states and O-based states of the
same symmetry. If the O bands are well below the Ni levels (for a
large, positive charge transfer energy compared to $\abs{t_{pd}}$),
then these hybridized states are predominantly located on the Ni; this
is the CD scenario. On the other hand, if the charge transfer energy
is very negative, then the hybridized states will live primarily on
the O; this corresponds to the NCT scenario. The reality is likely to
be somewhere in between, where the probability to be on the Ni is
neither 1 nor 0. Our
effective model describes this entire continuum of possibilities, for
appropriate choices of the parameters.

%
%

\section{Hartree-Fock calculation}\label{sec:HF-calc}

We study the Hamiltonian of Eq. (\ref{Ham}) within the Hartree-Fock
approximation. As usual, this implies finding the global minimum of
the average energy
\begin{equation}
\label{hf1}
E(\{ u \}) = \langle \Psi_{e}| \hat{H}(\{ u \})| \Psi_e\rangle
\end{equation}
where the Slater determinant $\ket{\Psi_e}$ describes the electronic
part, and the set $\{ u \} = (u_1, u_2,\dots)$ characterizes the
semiclassical distortions of the lattice.

\subsection{Lattice contributions}

First, we minimize the energy with respect to the lattice distortions $u_i$. This can be done easily using the Hellmann-Feynman theorem: \cite{Hellmann2015, Feynman1939}
\begin{equation}
\label{hf2}
\frac{d}{d u_j} E(\{ u \}) 
= \bra{\Psi_{e}}  \frac{\partial \hat{H}(\{ u \})}{\partial u_j} \ket{\Psi_{e}}.
\end{equation}
Note that Eq. ({\ref{hf2}) holds despite the fact that  $\ket{\Psi_{e}}$ is not an eigenstate of $\hat{H}(\{ u \})$ (as assumed in the usual proof \cite{Griffiths2005}). This is because a stronger proof is available, one that shows the theorem to hold for any sufficiently optimized variational state, not just for exact eigenstates \cite{Jensen2007} (in our case, we also confirmed this numerically). An optimal variational state $\ket{\Psi_e}$ satisfies the stationarity condition:  $\frac{\delta E}{\delta\Psi_e(u)} = 0$ (this is a shorthand notation replacing all the derivatives with respect to all the one-particle orbitals defining $|\Psi_e\rangle$). This condition justifies why the second term vanishes in the identity $\frac{d}{d u_j}E(\{ u \}) = \frac{\partial E\{ u \})}{\partial u_j} + \frac{\delta E\{ u \}}{\delta\Psi_e(u)} \frac{\partial \Psi_e(u)}{\partial u_j}$, which then leads to Eq. ({\ref{hf2}).

Thus, for our Hamiltonian we obtain the minimization condition, at each site $j$:
\begin{equation}
\label{hf3}
u_j + \alpha u_j^3 = \langle \hat{n}_j \rangle - 1,
\end{equation}
where from now on we use the short-hand notation $\langle\hat{O} \rangle = \langle \Psi_e | \hat{O} |\Psi_e \rangle$ for any electronic operator.

To make further progress, we use the experimentally well-established fact \cite{Garcia-Munoz1992a} that the octahedra alternate between expanded and collapsed ones, so that $u_j = u e^{i\mathbf{Q}_c\cdot\mathbf{R}_j}$, where $\mathbf{Q}_c = \pi(1,1,1)$. This immediately implies the appearance of a charge modulation $\langle \hat{n}_j \rangle = 1 + \delta e^{i\mathbf{Q}_c\cdot\mathbf{R}_j}$, where the amplitude of the lattice distortion $u$ is directly linked to the amplitude of the charge modulation $\delta$ by:
\begin{equation}\label{eq:latt-self-consis}
u + \alpha u^3 = \delta
\end{equation}
This equation shows that in our model, the existence of a charge
modulation $\delta\ne 0$ forces the appearance of a lattice distortion $u \ne
0$, and \textit{vice versa}. This depressed\footnote{\textit{i.e.} a cubic equation lacking a quadratic term.} cubic equation admits the
exact solution using Cardano's formula (see Appendix \ref{app:latt-self-consis} for details):
\begin{equation}
\label{hf5}
u = \delta \frac{3}{2\beta^{\frac{1}{3}}} \left[  \left( \sqrt{ 1 + \frac{1}{\beta} }+ 1 \right)^{\frac{1}{3}} - \left( \sqrt{ 1 + \frac{1}{\beta}  }  - 1 \right)^{\frac{1}{3}}  \right],
\end{equation}
with $\beta = \frac{27}{4} \alpha \delta^2.$

It is useful to consider this expression in some limiting cases:  in
the case of vanishing anharmonicity, $\alpha \rightarrow 0$, already from Eq.
(\ref{eq:latt-self-consis}) we see that $u = \delta$. For a fixed value of $\delta$, increased anharmonicity $\alpha$ will lead to a decrease of $u$. Indeed, in the case of infinite
anharmonicity, $\beta \rightarrow \infty$ and we find
\begin{equation}
u  \approx \delta \frac{3}{2\beta^{\frac{1}{3}}} \left[ 2^{1/3} \right] \rightarrow 0.
\end{equation}

The exact solution listed in Eq. (\ref{hf5}) is very convenient
because it allows us to substitute for any $u$
dependence in the Hartree-Fock equations (discussed next), and have
them depend only on the electronic mean-fields.

\subsection{Electronic contributions} 

We follow the usual steps, briefly summarized here for completeness, to derive the Hartree-Fock (HF) equations. Any Slater determinant has the general form:
\[
\ket{\Psi_e} = \prod_p \hc{a}_p \ket{0},
\]
where the appropriate number of electrons (here equal to the number of
sites in our lattice) are created. The new states and old states are
related by a unitary transformation
\[
\hc{d}_{ia\sigma} = \sum_n \phi_n^*(ia\sigma) \hc{a}_n.
\]
The goal is to determine the optimal $\phi_n(ia\sigma)$ which minimize the
total energy $E(\{ u\})= \langle \hat{H}(\{ u\})\rangle$. The evaluation of this
expectation value, and its minimization with respect to all properly
normalized $\phi_n(ia\sigma)$ proceeds in the usual way. As always, the
resulting HF equations depend on various mean-field expectation values
$\langle \hc{d}_{ia\sigma} d_{ib \sigma'} \rangle$ (because all interactions are local,
terms with $i \ne j$ do not appear).

We constrain these mean-fields to have the most general forms
consistent with the 4-site unit cell found experimentally in the
magnetically ordered state. Specifically, we set:
\begin{widetext}
  \begin{align}
& \langle \hc{d}_{ia\sigma} d_{ia\sigma} \rangle = \frac{1}{4} \left[ 1 + \delta
      e^{i\mathbf{Q}_c\cdot \mathbf{R}_i} \right] + \frac{\sigma}{2} \Big[
      S_{\text{FM}} + S_{\text{AFM}}e^{i\mathbf{Q}_c\cdot \mathbf{R}_i} +
      S_{\text{1z}} \cos(\mathbf{Q}_m\cdot \mathbf{R}_i) + S_{\text{2z}}
      \sin(\mathbf{Q}_m\cdot \mathbf{R}_i) \Big], \label{hfe1}\\ &\langle
    \hc{d}_{ia\sigma} d_{ia\bar{\sigma}} \rangle = \frac{1}{2} \left[ S_{\text{1x}}
      \cos(\mathbf{Q}_m\cdot \mathbf{R}_i) + S_{\text{2x}}
      \sin(\mathbf{Q}_m\cdot \mathbf{R}_i) \right], \label{hfe2}\\ & \langle
    \hc{d}_{ia\sigma} d_{i\bar{a}\sigma} \rangle = O_1 + O_2 e^{i\mathbf{Q}_c\cdot
      \mathbf{R}_i} + \sigma \Big[ Z_1 + Z_2 e^{i\mathbf{Q}_c\cdot
        \mathbf{R}_i} + Z_3 \cos(\mathbf{Q}_m\cdot \mathbf{R}_i) + Z_4
      \sin(\mathbf{Q}_m\cdot \mathbf{R}_i) \Big], \label{hfe3}\\ &\langle
    \hc{d}_{ia\sigma} d_{i\bar{a}\bar{\sigma}} \rangle = X_1 \cos(\mathbf{Q}_m\cdot
    \mathbf{R}_i) + X_2 \sin(\mathbf{Q}_m\cdot \mathbf{R}_i).\label{hfe4}
\end{align}     
\end{widetext}
Equation (\ref{hfe1}) is consistent with the condition that $\langle
\hat{n}_i\rangle = \sum_{a,\sigma} \langle \hc{d}_{ia\sigma} d_{ia\sigma} \rangle = 1 + \delta
e^{i\mathbf{Q}_c\cdot \mathbf{R}_i}$. The other terms in it allow for
various possible magnetic orders with a
non-vanishing $z$-axis spin expectation value:
 \begin{align}
\langle \hat{S}_{i,z}\rangle =& \frac{1}{2} \sum_{a,\sigma} \sigma \langle \hc{d}_{ia\sigma} d_{ia\sigma} \rangle =
S_{\text{FM}} + S_{\text{AFM}}e^{i\mathbf{Q}_c\cdot \mathbf{R}_i} \nonumber \\ +
&S_{\text{1z}} \cos(\mathbf{Q}_m\cdot \mathbf{R}_i) 
+ S_{\text{2z}} \sin(\mathbf{Q}_m\cdot \mathbf{R}_i).
\end{align}     
 If only $S_{\text{FM}} \ne 0$, the order is ferromagnetic (FM), or $\uparrow \uparrow \uparrow \uparrow
 $ (for simplicity, we only show the order inside one 4-site unit
 cell); if only $S_{\text{AFM}}\ne 0$, the order is antiferromagnetic 
 (AFM) $\uparrow \downarrow \uparrow \downarrow $; finally, having $S_{\text{1z}} \ne 0$ or
 $S_{\text{2z}} \ne 0$ further breaks translational symmetry, resulting
 in states with order like $\uparrow \!\!0 \!\! \downarrow \!\! 0$ and $0 \!\! \uparrow
 \!\!0\!\! \downarrow$, respectively. Combinations of two or more finite
 expectation values lead to yet more possibilities, for example having
 both $S_{\text{FM}} \ne 0, S_{\text{AFM}}\ne 0$ implies a ferrimagnetic
 order $\Uparrow \downarrow \Uparrow \downarrow$ or $\Uparrow \uparrow \Uparrow \uparrow$, depending on the relative magnitude of the parameters, \textit{etc}.
 
Equation (\ref{hfe2}) allows for non-collinear magnetic orders,
because a finite $S_{\text{1x}}$ and/or $S_{\text{2x}}$ imply
non-vanishing $\langle \hat{S}_{i,x}\rangle$. In particular, a solution with finite
$S_{\text{1z}} = S_{\text{2x}}$ while all other values are set to zero
implies the order $\uparrow \rightarrow \downarrow \leftarrow$, which is one of the possible
candidates. To conclude, Eqs. (\ref{hfe1}) and (\ref{hfe2}) allow the
realization of any magnetic order with a 4-site unit cell consistent with the
ordering vector $\mathbf{Q}_m=\mathbf{Q}_c/2$.

Similarly, Eqs. (\ref{hfe3}) and (\ref{hfe4}) allow for various
orbital and magnetic+orbital orders, respectively. To the best of our
knowledge, there is no experimental signature of any orbital ordering in the rare-earth nickelates,
\cite{Scagnoli2006} and numerical calculations suggest that orbital
order states are expected to be relatively high energy.
\cite{Mizokawa1996} Previous work \cite{Peters2009a} 
found orbitally ordered ground states for $U'/W \approx (U - 2J)/6 > 2$. Here, $W \sim
6$ is the bandwidth of the non-interacting system for $t_1 = 1$ and
$t_2, t_4 < 0.3$. For a realistic $J/t_1 \sim 1-2$, this implies $U/t_1 > 14-16$, which is enough to stabilize a conventional spin and orbital ordered Mott phase with no charge disproportionation. Our model is consistent with this, in the sense that the
lowest energy self-consistent states for experimentally reasonable
values of the charge modulation $\delta$ always have vanishing $O,X,Z$ values. However,
we have also found regions of parameter space where orbital ordered states
appear to have the lowest energy, as discussed below.
For convenience, in the following we will still set $ \langle\hc{d}_{ia\sigma}
d_{i\bar{a}\sigma'} \rangle = 0$ so as to keep the equations shorter, with the
full form available in the appendix. We emphasize that in the regions of
parameter space relevant to us, i.e. the neighborhood of the region where
$\delta \sim 0.3$-$0.4$ in agreement with experimental observations, we have tested explicitly that the ground
states do not exhibit orbital order, by running self-consistency loops
where these mean-fields $O, X, Z$ had finite initial values. The resulting
self-consistent solutions either converged to vanishing values for
these orbital mean-fields, or had much higher total energy than
self-consistent states without orbital order, in accord with earlier
findings. \cite{Mizokawa1996}

The corresponding HF equations are identical to those arising from a
non-interacting Hamiltonian $\hat{H}_{\rm eff}$ (which can be thought
of as being the properly-factorized counterpart of the original
$\hat{H}$):
\begin{widetext}
	\begin{eqnarray}\label{eq:big-guy}
H_{\text{eff}} = &&\sum_{\mathbf{k}ab\sigma} t_{ab}(\mathbf{k})
\hc{c}_{\mathbf{k}a\sigma} c_{\mathbf{k}b\sigma} + \sum_{\mathbf{k}a\sigma} \Big[
  \frac{3U-5J}{4} - \frac{\sigma}{2} (U+J) S_{\text{FM}} \Big]
\hc{c}_{\mathbf{k}a\sigma} c_{\mathbf{k}a\sigma} \nonumber\\ &&+
\sum_{\mathbf{k}a\sigma} \Big[ \frac{3U-5J}{4}\delta - 2\epsilon_b u - \frac{\sigma}{2} (U+J)
  S_{\text{AFM}} \Big] \hc{c}_{\mathbf{k}+\mathbf{Q}_c,a\sigma}
c_{\mathbf{k}a\sigma} \nonumber\\ &&-\frac{\sigma}{4}(U+J) \sum_{\mathbf{k}a\sigma}
\Big[ (S_{\text{1z}} - i S_{\text{2z}})
  \hc{c}_{\mathbf{k}+\mathbf{Q}_m,a\sigma} c_{\mathbf{k}a\sigma} +
  (S_{\text{1z}} + iS_{\text{2z}}) \hc{c}_{\mathbf{k}-\mathbf{Q}_m,a\sigma}
  c_{\mathbf{k}a\sigma} \Big] \nonumber\\ &&-\frac{U+J}{4}
\sum_{\mathbf{k}a\sigma}\Big[ (S_{\text{1x}} - i S_{\text{2x}})
  \hc{c}_{\mathbf{k}+\mathbf{Q}_m,a\sigma} c_{\mathbf{k}a\bar{\sigma}} +
  (S_{\text{1x}} + iS_{\text{2x}}) \hc{c}_{\mathbf{k}-\mathbf{Q}_m,a\sigma}
  c_{\mathbf{k}a\bar{\sigma}} \Big].
\end{eqnarray}
\end{widetext}
This Hamiltonian has four more similar lines of terms involving
various $O,Z,X$ mean-fields which we do not write here explicitly, as
discussed above. We remind the reader that the lattice parameter $u$ appearing in the
second line is given by Eq. (\ref{hf5}).

After Fourier transforming,  $ H_{\text{eff}} =
\sum_{\mathbf{k}}^{} \hc{\psi}_{\mathbf{k}} h(\mathbf{k}) \psi_{\mathbf{k}}$, where $
\hc{\psi}_{\mathbf{k}}=
(\hc{\psi}_{\mathbf{k}z\uparrow},\hc{\psi}_{\mathbf{k}\bar{z}\uparrow},
\hc{\psi}_{\mathbf{k}z\downarrow}, \hc{\psi}_{\mathbf{k}\bar{z}\downarrow})$ and
$\hc{\psi}_{\mathbf{k}a\sigma} = (\hc{c}_{\mathbf{k}a\sigma},
\hc{c}_{\mathbf{k}+\mathbf{Q}_m,a\sigma},
\hc{c}_{\mathbf{k}+\mathbf{Q}_c,a\sigma},
\hc{c}_{\mathbf{k}-\mathbf{Q}_m,a\sigma})$. The $16\times16$ matrix
$h(\mathbf{k})$ can be directly read from Eq. (\ref{eq:big-guy}) and is
trivial to diagonalize numerically.

The self-consistent HF ground state is then straightforward to find,
at least in principle. We start with an initial guess for the
mean-field parameters $\mathbf{w}^{(0)} = (\delta^{(0)},
S_{\text{FM}}^{(0)}, S_{\text{AFM}}^{(0)}, ...)$. These can be chosen
either so as to test if a certain state, \textit{e.g.} $\uparrow \rightarrow
\downarrow \leftarrow$, is self-consistent, or by choosing random values for all
these fields. We have always checked all the ``simple'' magnetic orders to
see if they are self-consistent and if yes, what is their
corresponding energy. However, in all cases we have also run a
multitude of searches starting with random initial conditions,
to make sure we are not missing a better candidate.

Once the mean-field parameters are chosen, the Hamiltonian
(\ref{eq:big-guy}) is diagonalized at all allowed $\mathbf{k}$-points in
the Brillouin zone, and its ground state at quarter-filling is
identified. Using it, we compute the new mean-field parameters
$\mathbf{v}^{(0)}= (\delta^{(0)}, S_{\text{FM}}^{(0)},
S_{\text{AFM}}^{(0)}, ...)$ based on Eqs. (\ref{hfe1})-(\ref{hfe4}).
For example, $\delta^{(0)} = \frac{1}{N} \sum_{ia\sigma} e^{i\mathbf{Q}_c \cdot
  \mathbf{R}_i}\langle \hat{n}_{ia\sigma} \rangle = \frac{1}{N} \sum_{\mathbf{k} a \sigma} \langle \hc{c}_{\mathbf{k}+\phi_c,a\sigma} c_{\mathbf{k}a\sigma} \rangle$, \textit{etc}. We then compute the residual
$\epsilon^{(0)} = \left| \mathbf{v}^{(0)} - \mathbf{w}^{(0)} \right|^2$. If
this is below the desired accuracy, then convergence has been reached.
If not, we set new values for the mean-field parameters:
$\mathbf{w}^{(1)} = \mathbf{v}^{(0)}$ (this is a simplification: in reality we use a better
choice, discussed below), and iterate until either the desired
accuracy is reached, or the maximum iteration count is surpassed and
this search is abandoned.

Once self-consistency is reached, \textit{i.e.} $\mathbf{w}\approx\mathbf{v} $,
the total energy associated with the set $\mathbf{w}$ of mean-field
parameters is given by:
\begin{equation}
\label{hf8}
\frac{E}{N} = \frac{1}{N} \sum_p E_p - \frac{1}{N} \langle \hat{H}_{e-e} \rangle + 2\epsilon_b
\left(\frac{u^2}{2} + \frac{au^4}{4} \right).
\end{equation}
Here $E_p$ are the eigenenergies of the occupied states, and
\begin{align}
\langle \hat{H}_{e-e} \rangle =& \frac{3U-5J}{8} (1+\delta^2) - \frac{U+J}{2} \Big(
S_{\text{FM}}^2 \nonumber\\ & + S_{\text{AFM}}^2 +
\frac{S_{\text{1x}}^2 + S_{\text{1z}}^2 + S_{\text{2x}}^2 +
  S_{\text{2z}}^2}{2} \Big)
\end{align}
(again, terms proportional to the $O, Z, X$ fields are omitted here and are instead given in the appendix).

After multiple searches for various initial conditions, the set
$\mathbf{w}\approx\mathbf{v} $ corresponding to the lowest total energy $E$
is declared as the HF ground state, and its magnetic (and charge, orbital, lattice...) order is read off
from its mean-field parameters.

While all this seems straightforward, in reality the problem is
complicated by the large number of mean-field parameters whose
convergence is sought: 15 when the $O,Z,X$ fields are included
explicitly, and 7 otherwise. Searching for a local minimum in this many-dimensional space
is non-trivial, and for too simplistic update rules such as $
\mathbf{w}^{(n+1)}= \mathbf{v}^{(n)}$, it can take extremely many
iteration steps before self-consistency is reached, if it is reached at all.

Because of this, we briefly mention here a few strategies that we found very useful:

(a) A much better update is an interpolation of the type:
\begin{equation}\label{simple-update}
\mathbf{w}^{(n+1)} = \alpha \mathbf{v}^{(n)}+ (1 - \alpha)\mathbf{w}^{(n)}.
\end{equation}
The literature on nonlinear iterative equation solution techniques
suggests that the choice of the mixing parameter $\alpha$ is typically
problem-specific. \cite{Banerjee2016} We find the smoothest and most
reliable convergence over most of the parameter space of interest
occurs for $\alpha = 0.3$. This type of update can avoid the iteration
being stuck in a loop, or helplessly hopping on either side of a
``flat minimum''.

(b) However, we found that this approach works best in conjunction
with Pulay mixing, or direct inversion in the iterative subspace
(DIIS). The idea behind DIIS, originally developed for high-parameter
Hartree-Fock quantum chemistry calculations by Pulay, \cite{Pulay1980}
is as follows: suppose that a sequence of solutions ${
  \mathbf{v}^{(n+1)}, \mathbf{v}^{(n+2)}, ... \mathbf{v}^{(n+N)}}$ to
the nonlinear system has been generated, from some initial solution
step $n$. Together, they span a linear subspace $W_{N,n} = \spn\{
\mathbf{v}^{(n+1)}, \mathbf{v}^{(n+2)}, ... \mathbf{v}^{(n+N)}\}$
within the higher-dimensional (possibly nonlinear) parameter manifold
$W$. It is then possible to pick the ``best possible'' vector
$\mathbf{w}^{(N)}$ within this linear subspace, by minimizing the
error $ \epsilon = ||\sum_i^N \mathbf{v}^{(i)} - \mathbf{w}^{(i)}
||^2$, which amounts to solving an $N+1\times N+1$ system of linear
equations (hence the ``inversion of the iterative subspace'' -- for
more details, see Ref. [\onlinecite{Pulay1980}]).

The DIIS can quickly maximize the potential of the vectors within the
subspace $W_{N,n}$, but if the true solution is outside the subspace
by more than the allowed residual $\epsilon_0$, then no matter how
many times the Pulay mixing is carried out, it will not result in
improved convergence. The solution \cite{Banerjee2016} is to
intersperse Pulay mixing with regular updates of the form
(\ref{simple-update}), with a given periodicity $k$ (typically $ k =
3$). The algorithm thus alternates between expanding its iterative
subspace, and finding the lowest residual vector within it, resulting
in optimal convergence for most parameter values. Our experience is
that it can even arrest divergences from round-off error accumulation.
Often even small errors (say, in mean-fields that should be zero for a
particular type of converged state) can lead to rapid divergence of
the HF iterative trajectory away from a self-consistent point. Yet
DIIS appears to nullify that tendency, firmly slashing those creeping
mean-field magnitudes back to zero and guiding the trajectory towards
the self-consistent point.

(c) {Parallelization}: While the iterative loop is not easy to
parallelize efficiently, there are higher-level parallelization
opportunities: (i) multiple initial guesses can be iterated in
parallel, for a given set of parameter values $U,J, \epsilon_b, t_1, \dots$; and
(ii) ground states can be found at multiple parameter values, in
parallel. We opted for option (ii) due to ease of implementation and
data management, together with linear speed-up of the calculation
(because ground states corresponding to different parameter values are
independent from each other, there is no overhead to the
parallelization).

(d) {Boot-strapping:} While calculations at neighbouring points in the
parameter space are independent from each other, we expect small
changes in the values of the various parameters $U, J, \epsilon_b, t_1, ...$
to normally lead to small changes in the nature of the ground state
$\mathbf{w}$. As such, an already converged solution from a
neighbouring point can result in fast convergence to a similar kind of
converged solution at the current point. Where available, we
included this option in the set of initial mean-field guesses.

Finally, we found that there are two key parameters that need to be
tested for convergence: the cutoff iteration count
$n_{\text{max-iter}}$, and the number of points sampled in the
Brillouin zone $n_{\text{num-k-pts}}=N^3$. The sampling of the
momentum space slows down the calculation time $\sim
\mathcal{O}(N^3)$, so choosing too large an $N$ is costly while too
low a value introduced finite-size effects. Similarly, the cutoff
iteration count needs to be large enough to allow convergence for the
interesting solutions, but not so large as to waste computation time
on ``dead-end'' iterative trajectories that never converge. We found
$N=25$ and $n_{\text{max-iter}} = 500$ to be optimal for our
purposes, taming the error in mean-fields to below $\epsilon_0 =
10^{-4}$ and in ground state energies to less than $10^{-3}$.

In total, to obtain a typical phase diagram, we carry out anywhere from 400 to 1,600 calculations (depending on the desired resolution), each of which starts from a pool of 20-30 initial guesses and proceeds through anywhere from 5 to 500 iterative steps (where each iterative step involves the diagonalization of roughly 15,000 $16\times16$ subblocks of the Hartree-Fock matrix, as well as the calculation of the density matrix and the order parameters $\mathbf{w}$).

\begin{figure}[b]
	\includegraphics[width=0.9\columnwidth]{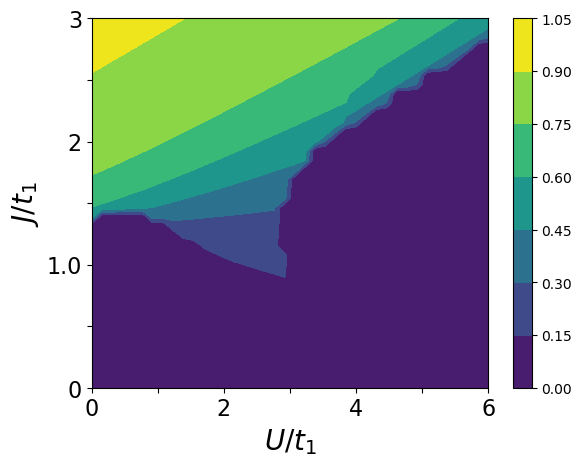}
	\caption{ Charge modulation $\delta$ (see colour scale) in the HF ground-state in the $U$-$J$ plane. Other parameters are $t_1 = 1, t_2 = 0.15, t_4=0, \epsilon_b=0$. Resolution is $40\times40$.  }
	\label{fig:chrdisp-U-J}
\end{figure}

\section{Results}\label{sec:res-and-disc}

\begin{figure*}
	\includegraphics[width=0.95\textwidth]{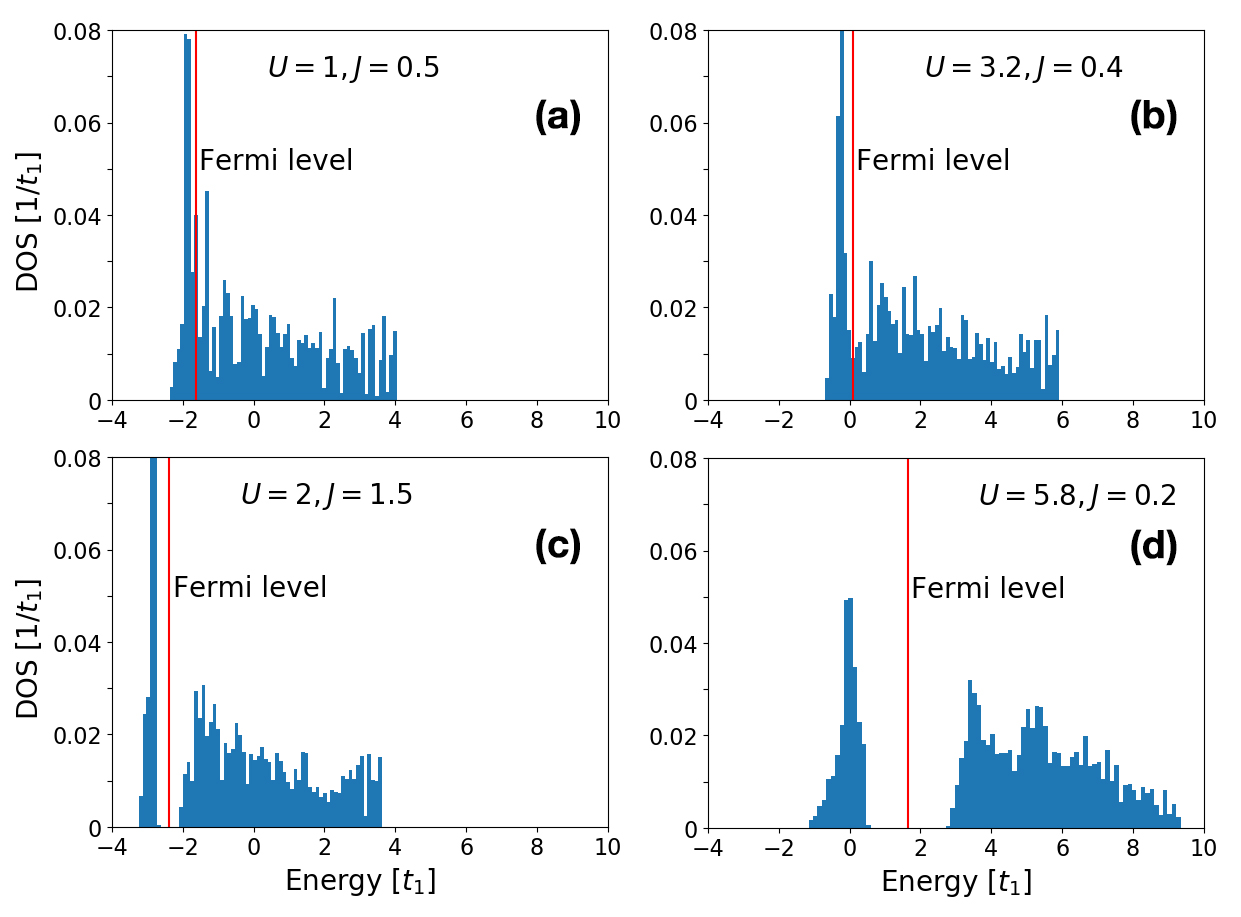}
	\caption{ Representative density of states for the various phases in Fig. \ref{fig:chrdisp-U-J}: (a) metallic phase, $U = 1.077, J = 0.538$; (b) itinerant magnetic phase (see the magnetic phase analysis below for more details about this phase), $U = 3.231, J = 0.385$; (c) charge modulated phase, with a clear gap at the Fermi level, $U = 2.0, J = 1.538$; (d) Mott insulating phase, $U = 5.846, J = 0.231$. Other parameters as in Fig. \ref{fig:chrdisp-U-J}. Notice the van Hove singularity at the lower band edge in all the diagrams: its presence is due to the nonzero $t_2$ parameter, which introduces a strong asymmetry to the DOS. More on this below. }
	\label{fig:DOS-met/ins}
\end{figure*}

The results shown here focus on areas of the phase diagram with
essentially no orbital order. As mentioned, our searches for converged
self-consistent states with non-trivial orbital order never produced a
viable candidate for an HF ground state in the region of the parameter
space with finite charge modulation $\delta \sim 0.3$-$0.4$. However, we did
find that sometimes orbitally ordered states were indeed the ground
state configuration in other regions of the phase diagram, namely towards the large U Mott limit. Most notably,
the antiferromagnetic ferroorbital order, $\uparrow\downarrow\uparrow\downarrow +  XxXx$ (marked ``afm-ferroorb'' in the phase diagrams below), and other orbitally ordered possibilities
(marked miscellaneous, or ``misc'').
Given that such orders are found outside the experimentally relevant
parameter regime, from now on we focus only on the HF parameters $\delta,
S_{\text{FM}}, S_{\text{AFM}}, S_{\text{1z/x}}, S_{\text{2z/x}}$ and
leave the investigation of the regions with stable orbital order to
future work. The lattice distortion $u$ is related to $\delta$ through Eq.
(\ref{hf5}).

To get a basic idea of the dependence of the charge modulation
$\delta$ on the electronic interaction strengths $U$ and $J$, we start
by showing in Fig. \ref{fig:chrdisp-U-J} a contour plot of the charge
modulation in the $U$-$J$ plane, for $t_1 = 1, t_2 = 0.15, t_4=0,
\epsilon_b = 0$. We set $t_1 = 1$ throughout this paper, so all energy scales are reported in  units of $t_1$. Our results
are in qualitative agreement with earlier findings at
similar parameter values. \cite{Lee2011} Unsurprisingly, for small
values of $U$ and $J$, we find $\delta = 0$ and the system is fully
metallic, with no gaps in the band structure, as can be seen from typical (volume) densities
of states (DOS) for the various phases, given in Fig.
\ref{fig:DOS-met/ins}. The vertical scale in the DOS plots is normed to the total volume of the crystal, i.e. $1/N^3$ (the lattice constant is set to 1). For large $J$ and small $U$,  in
the spirit of Hund's rule, the electrons find it preferable to occupy both orbitals
at the same site, leading to ever-increasing charge modulation between
neighboring sites and strong insulating behavior (this corresponds to the picture discussed in Ref. [\onlinecite{Mazin2007}]). In the other
extreme, for large Hubbard $U$ and small $J$, no charge modulation
arises because of the prohibitive cost of double occupancy. As a
result, the system remains itinerant up to fairly large values of $U$,
where it switches to a Mott insulator.

\begin{figure}[t]
	
	\includegraphics[width=0.48\textwidth]{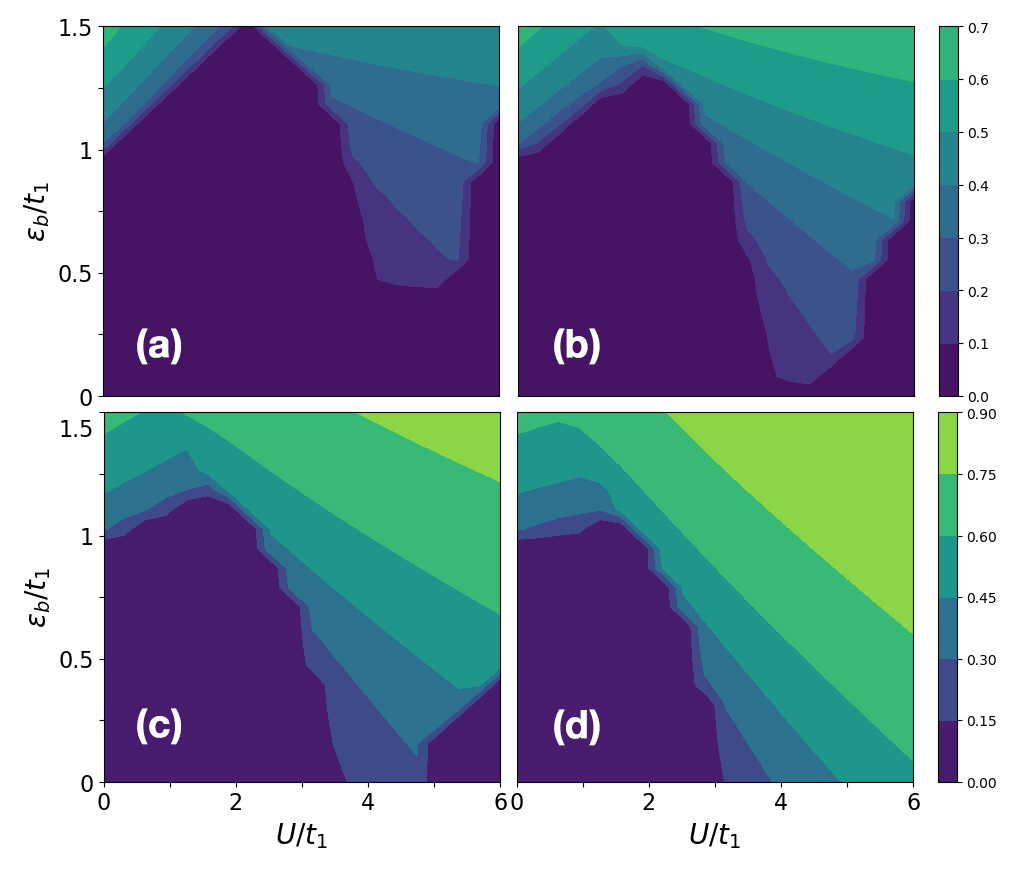}
	\caption{ Charge modulation $\delta$ (see color scale) in the HF ground-state, as a function of $U$ and $\epsilon_b$, for $J/U=0.2, 0.3$ ((a) and (b), respectively) and $0.4, 0.5$, ((c) and (d), respectively). Other parameters are $t_1 =1, t_2=0.15, t_4=0.25, \alpha=1$. Resolution is $20 \times 20$. }
	\label{fig:chrdisp}
\end{figure}

Of course, the interesting question is how charge modulation is
modified by the coupling to the lattice. To probe this, we fix various
ratios of $J/U=0.2, 0.3, 0.4, 0.5$ (they can be thought of as line
slices of the phase diagram in Fig. \ref{fig:chrdisp-U-J} emanating
from the origin) and tune the lattice coupling constant $\epsilon_b$,
while keeping fixed the values of the other parameters $t_1 = 1, t_2 = 0.15,
t_4 = 0.25, \alpha = 1$ (we added the 4$^{\text{th}}$ nearest
neighbor hopping to watch the lattice interact with all the
ingredients of the model -- explicit results on its influence on the
phase diagram will be discussed below). These results are shown in
Fig. \ref{fig:chrdisp}, while in Fig. \ref{fig:sloping} we indicate
whether the  HF ground state is metallic (blue) or
insulating (yellow).

Clearly, for small values of $U$ the system is metallic and
homogeneous, with $\delta=u=0$. Even in the absence of coupling to the
lattice, \textit{i.e.} when $\epsilon_b=0$, with increasing $J$ there
is a transition to a state with a finite charge modulation $\delta$,
which eventually becomes insulating if $J$ is large enough. This MIT
occurs faster for larger  $J/U$ ratios, as the tendency for
two-orbital occupancy encouraged by $J$ grows faster than the drive
toward single-site occupancy coming from increased $U$. If the
coupling to the lattice is turned on, we find that $\delta$ increases
with $\epsilon_b$ everywhere, and the system is more
likely to become insulating: thus there is positive cooperation between $J$ and the lattice coupling $\epsilon_b$.

\begin{figure}[t]
	
	\includegraphics[width=0.48\textwidth]{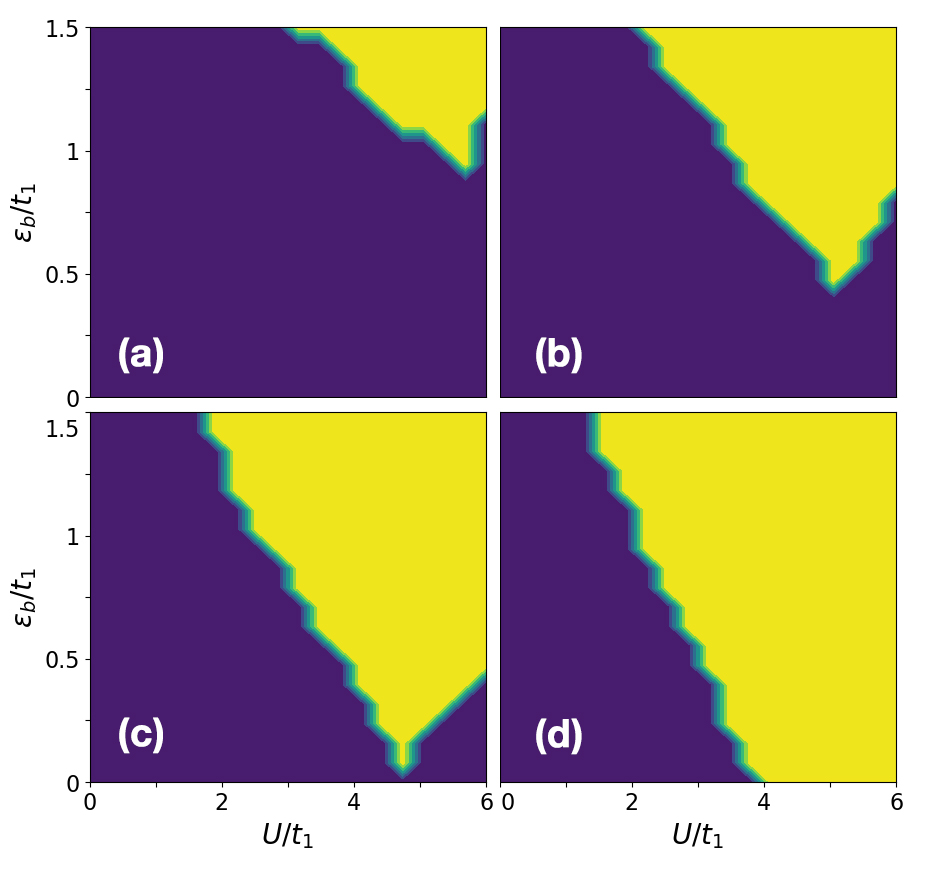}
	\caption{The HF ground-state is metallic (deep blue) or insulating (yellow). The results are shown in the $U$-$\epsilon_b$ space, for $J/U=0.2, 0.3, 0.4$ and 0.5, respectively (panels arranged as in Fig. \ref{fig:chrdisp}). All other parameters are as in Fig. \ref{fig:chrdisp}. Resolution is $20 \times 20$. }
	\label{fig:sloping}
\end{figure}

These results confirm the existence of the MIT, where the insulating state has a charge modulation $\delta \ne 0$ and a
lattice distortion $u \ne 0$. Coupling to the lattice increases the
likelihood of this insulating ground state with finite $u$, as expected.

\begin{figure}[t]
	\includegraphics[width=0.48\textwidth]{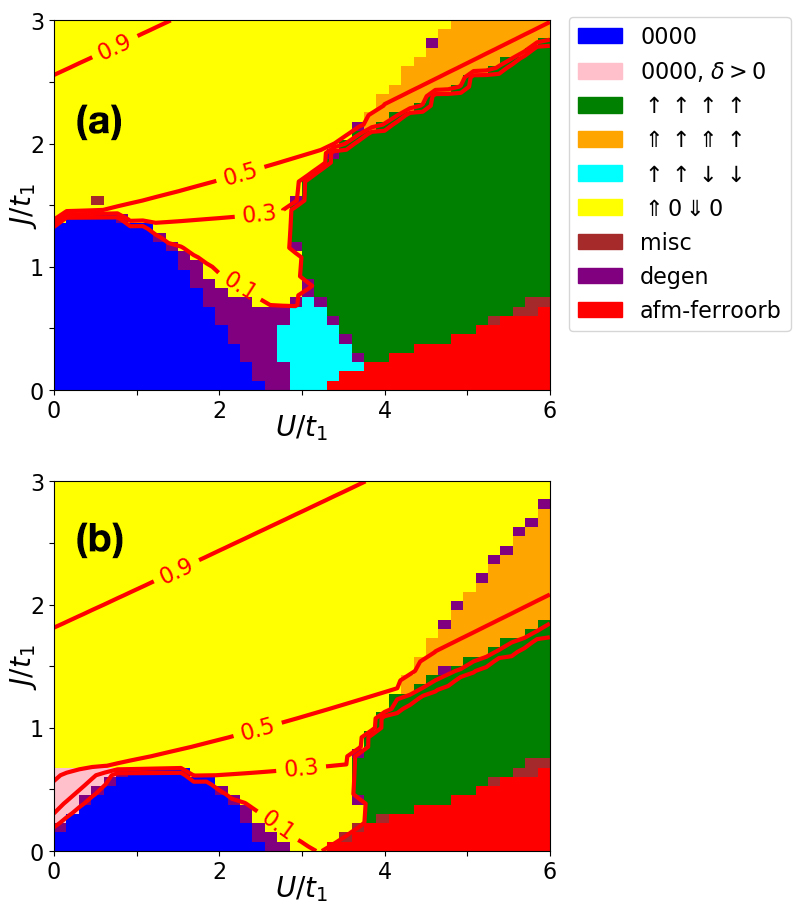}	
	\caption{ Top:  $U$-$J$ phase diagram of magnetic order, in the absence of coupling to the lattice ($\epsilon_b = 0$). Bottom: Same when coupling to the lattice is turned on ($\epsilon_b = 0.8$). Other parameters are $t_1 = 1, t_2 = 0.15, t_4 = 0, \alpha = 1$ for both. The black-line contours indicate the value of $\delta$. Resolution is 40 $\times 40$. }
	\label{fig:dominance}
\end{figure}

We now discuss the magnetic order found in the HF ground state,
and how it is influenced by the coupling to the lattice. Once again,
it is useful to start with the $U$-$J$ plane picture. In Fig.
\ref{fig:dominance} we show the magnetic order found in the HF ground
state for various values of $U$ and $J$ both in the absence and in the
presence of coupling to the lattice ($\epsilon_b=0$ and $\epsilon_b =
0.8$, respectively). The other parameters are kept fixed. The black contours
indicate the corresponding value of $\delta$. If we fix $\delta$ at
experimentally relevant values $\delta\approx 0.3-0.4$, we see that as
a function of increasing $U$ (and $J$ adjusting accordingly), the
system evolves through a variety of states. While charge modulation of
the experimentally appropriate magnitude appears to originate just
past the boundary of the metallic and insulating regimes at low $U$
and intermediate $J$, as we follow the contour line it quickly enters
the bulk of the $\Uparrow 0 \Downarrow 0$ magnetic phase already at $U
\approx 1$, before again falling on a phase line (this time between
the ferromagnetic and aligned-ferrimagnetic phases) around $U \approx
4$ for what seems like the rest of the contour. The fact that the
state $\Uparrow 0 \Downarrow 0$ is favored by larger $J$ and lies,
for the most part, above these $\delta$ values, speaks in its favor
as the preferred magnetic ground state. In addition, for a
decently-sized range of $U$ and $J$ values ($U$ from 1 to 3 and $J$
from 1 to 1.5 -- a range potentially consistent with experimental
values for these parameters) the contour remains squarely in the
$\Uparrow 0 \Downarrow 0$ phase, thus suggesting that this might be
the most energetically favorable 4-site spin alignment. Our phase
diagram agrees qualitatively with that of Ref. [\onlinecite{Lee2011}], although
there are sizable quantitative differences, which we attribute to the
different form of the on-site electronic Hamiltonian used.

We note that in between the magnetic phases there are often small
regions wherein the energy of several different magnetic orders are
indistinguishable to within our computational accuracy: such regions
are designated ``degenerate''. We emphasize that, compared to the
picture of Ref. [\onlinecite{Lee2011}], these regions do not correspond
to partially-disproportionated states of type $\Uparrow \uparrow\Downarrow\downarrow$ (in fact, such
states are never convergent within our model: see below): instead, the
states from neighboring phases (like, say, $\Uparrow 0 \Downarrow 0$ and $\uparrow \uparrow \downarrow \downarrow$ and
$\uparrow \rightarrow \downarrow \leftarrow$ in the lower middle of the phase diagram) all have roughly
the same energy, down to $\Delta E \sim 10^{-3}$.

When the coupling to the lattice is turned on, it acts to strongly
reduce the energy of the charge modulated states, as one would expect.
Not only does it lead to a significant shift of the $\Uparrow 0
\Downarrow 0$ magnetic order boundary, further increasing its
likelihood to be the  magnetic ground state; it also shifts the
charge modulation contours, leading to an enhancement of the aligned
ferrimagnetic phase for large $U$ and $J$, as well as the decoupling
of the charge and magnetic order phase lines below $U = 1$, revealing
a sizable metallic phase with partial charge modulation.

\begin{figure*}
	\includegraphics[width=0.95\textwidth]{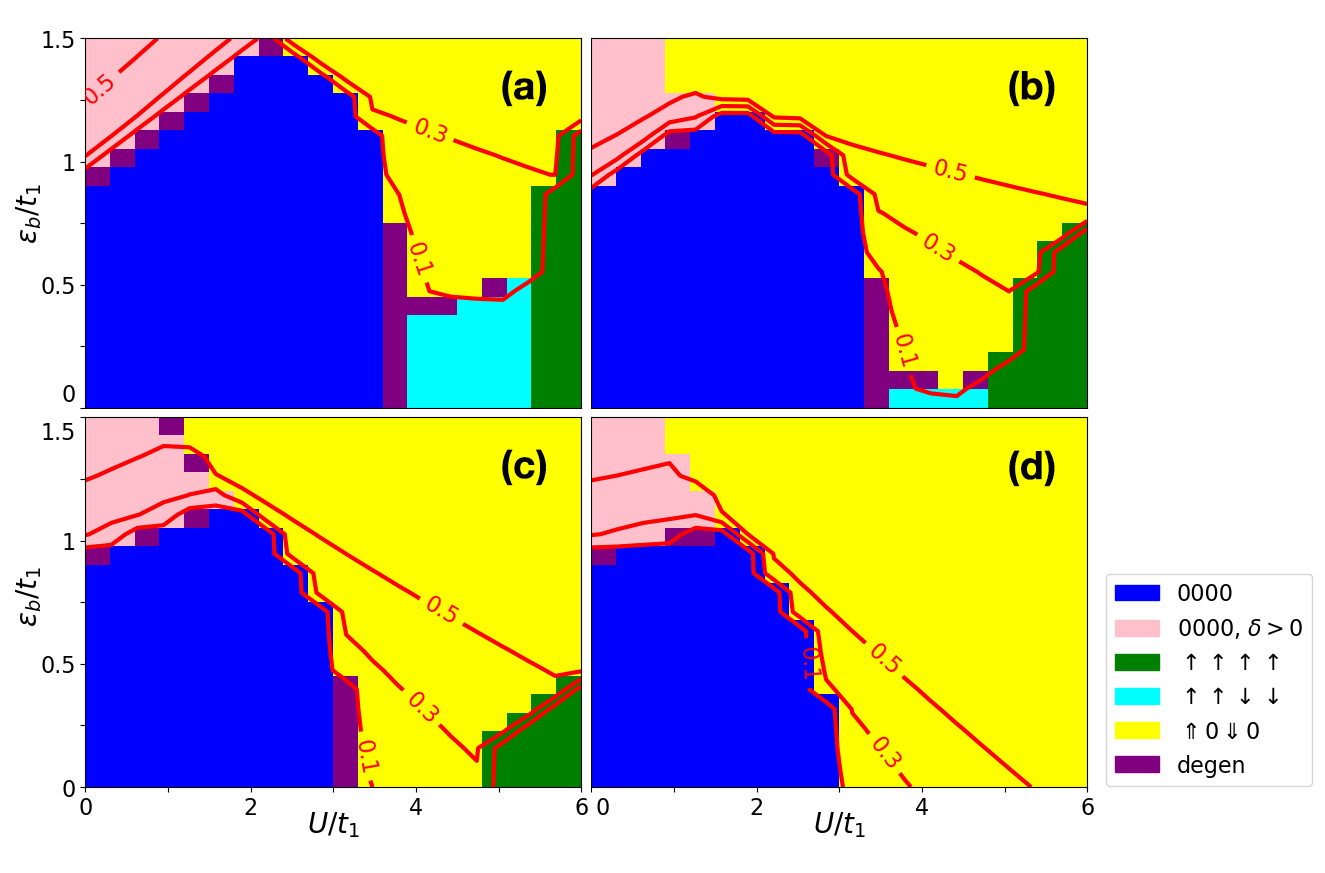}
	\caption{ Magnetic order in the HF ground state, as a function of $U$ and $\epsilon_b$, for $J/U=0.2, 0.3$ ((a) and (b), respectively) and $0.4, 0.5$, ((c) and (d), respectively). Other parameters are $t_1 =1, t_2=0.15, t_4=0.25, \alpha=1$. }
	\label{fig:magnetism}
\end{figure*}

Given this large-scale picture of the magnetic order, we once again fix
ratios $J/U$ and consider, in more detail, the impact of the lattice.
The data, shown in Fig. \ref{fig:magnetism}, demonstrates clearly that
an increase in coupling to the lattice leads to a creep of the
$\Uparrow 0 \Downarrow 0$ phase boundary: the stronger the coupling to
the lattice, the smaller the values of $U$ and $J$ that are required
to stabilize the fully charge modulated magnetic order. A curious
feature of this data, already noted in the previous paragraph, emerges
in the upper-left corner of the diagrams: while typically the contours
signaling the onset of charge modulation strongly follow the magnetic
ordering phase lines, we see that with increased coupling to the
lattice there is a decoupling of the onset of charge and magnetic
order. In other words, we obtain a charge modulated phase without any
magnetism. The lattice distortion acts to strongly gap out and flatten
the band structure: however, the Fermi level is still well within the
occupied band (a feature of the two-band Hubbard model), so the system
stays metallic and magnetic order does not arise. Thus the coupling to
the lattice acts to stabilize the charge order in the absence of
magnetism and localization. This is strongly reminiscent of the
decoupling of the MIT and the magnetic order transition for the Sm to
Y members of the nickelate series, a well-known feature of the
nickelates phase diagram, and could suggest that the strength of the
lattice coupling is ultimately responsible for determining whether the
charge and magnetic transitions are concurrent or not.

It is important to note that we usually find that many of the possible 4-site unit cell
magnetic orders turn out to be self-consistent within HF at the same values of parameters, and that
their HF energies can lie fairly close together. An example is shown
in Fig. \ref{fig:energy-scalings}, where the energy corresponding to
various self-consistent HF states is plotted vs. $\epsilon_b$, at fixed
values of the other parameters. This is, in essence, a slice along the $U \approx 4$ line of the top left panel in Fig. \ref{fig:magnetism}.  

\begin{figure}[b]
	
        \includegraphics[width=0.48\textwidth]{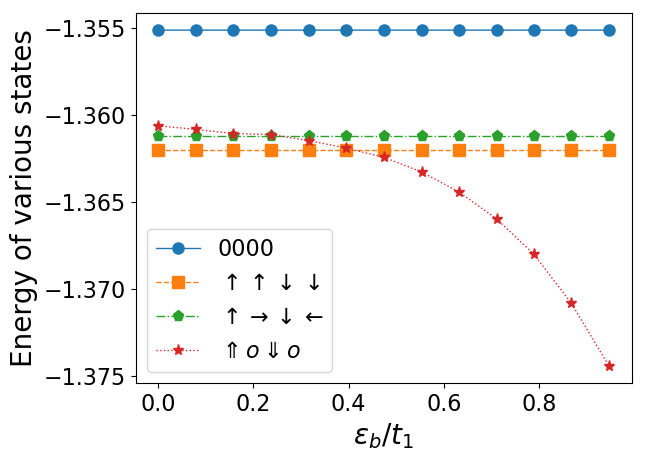}
	\caption{ Total energies of several converged self-consistent HF
          states as a function of $\epsilon_b$. The different colors
          correspond to different magnetic orders. The parameters are $U = 4.105, J = 0.2U, t_1 = 1, t_2 = 0.15, t_4 = 0.25, \alpha = 1$.}
	\label{fig:energy-scalings}
\end{figure}

At $\epsilon_b=0$, we find that the states
$\uparrow\rightarrow\downarrow\leftarrow$,
$\uparrow\uparrow\uparrow\uparrow$,
$\uparrow\uparrow\downarrow\downarrow$ and $\Uparrow 0 \Downarrow 0$
are all converged, with their HF energies per site being within $\sim
0.001t_1$ of each other. The $\uparrow \uparrow \downarrow \downarrow$
order has the lowest energy and thus is identified as the HF ground
state. The close spacing between these energies suggests that changing
any of the parameters and/or adding new ingredients -- in particular
coupling to the lattice -- may favor another magnetic order as the
ground state. Indeed, we see that as $\epsilon_b$ increases, the
energy of the $\Uparrow0 \Downarrow0$ state decreases and it
eventually becomes the new HF ground state. Note also that the
energies of the other magnetic states are independent of $\epsilon_b$
-- while this plot is made at a specific set of parameter values, this
pattern seems to hold across various parameter regimes of $U, J, t_i$.
There is a straightforward explanation to this: any state that
includes no charge modulation also includes no lattice distortion, as
$\delta \sim u$ through the lattice self-consistency equation (Eq.
\ref{hf5}). The real question is: why are the states with partial
charge modulation, \textit{e.g.} $\Uparrow \uparrow \Downarrow
\downarrow$ not convergent within this mean-field model? Such states
could potentially compete with $\Uparrow 0 \Downarrow 0$ order for
ground state status, as the coupling to the lattice is adjusted. We
find, however, that such states fail to converge no matter
the starting point, and in fact they are among the least stable, as
can be seen most clearly from cuts in the 15-dimensional parameter space that show iterative trajectories of the Hartree-Fock calculation (such diagrams are called ``Poincare sections'' in the dynamical systems literature). In Fig. \ref{fig:phase-portrait}, we plot
the evolution of the difference $S_{\text{1z}} - S_{\text{2z}}$ versus the charge
disproportionation $\delta$. Note that the states
$\uparrow\uparrow\downarrow\downarrow$ (central point) and $\Uparrow 0 \Downarrow 0$ (diagonal end points)
are well-defined  in this plane. Any solution not precisely on the
$S_{\text{1z}} - S_{\text{2z}} = 0$ line converges to $\Uparrow 0
\Downarrow 0$ instead.

\begin{figure}[b]
	
	\includegraphics[width=0.45\textwidth]{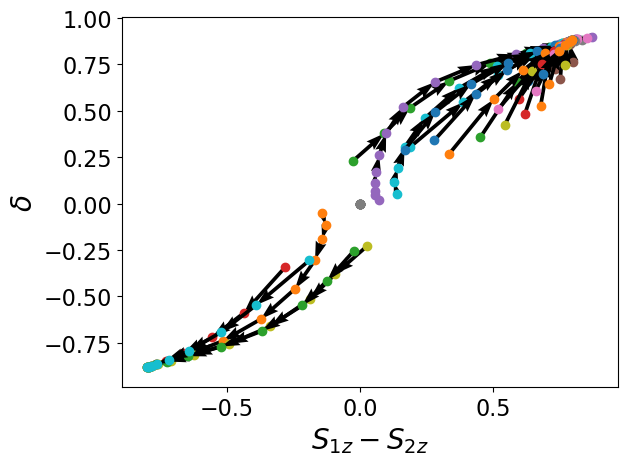}	
	\caption{ A phase portrait of the iterative sequences for various starting parameters in the $\delta, S_{1z}, S_{2z}$ parameter subspace. On the $x$ axis we plot the difference between the spin parameters of the sublattices, $S_{1z} - S_{2z}$: hence the perfect symmetry point is at the origin. On the $y$ axis is the disproportionation parameter $\delta$: the stable point $\Uparrow o \Downarrow o$ is thus at the top right. The parameters are $U = 5, J = 3, t_1 = 1, t_2 = 0.15, t_4 = 0.55, \epsilon_b = 0$. }
	\label{fig:phase-portrait}
\end{figure}

\begin{figure*}
	
	\includegraphics[width=0.99\textwidth]{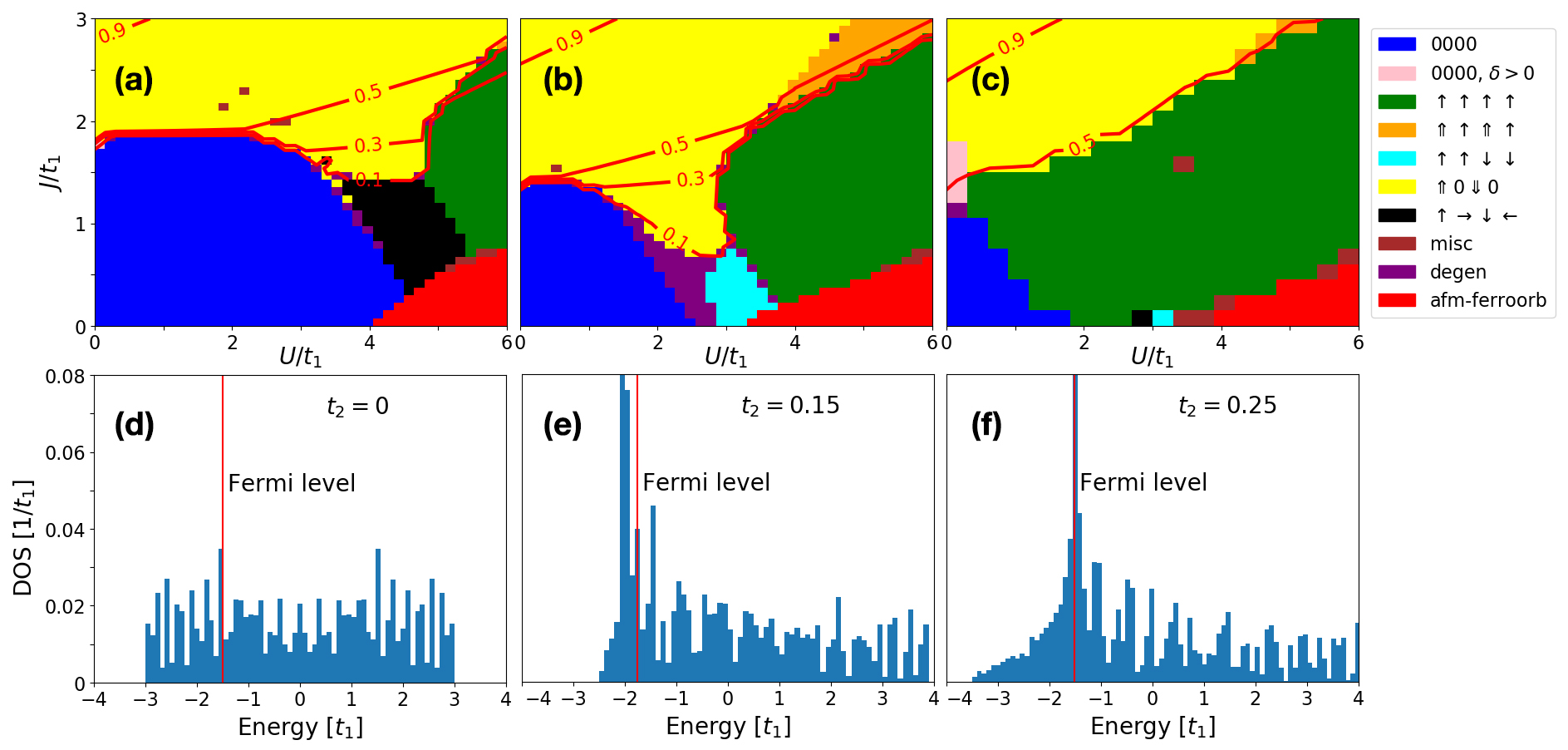}
	\caption{ Magnetic order in the HF ground state, as a function of $U$ and $J$, for $t_2 = 0, 0.15, 0.25$ ((a) (b) and (c), respectively). Other parameters are $t_1 =1, t_4 = 0, \epsilon_b = 0$. In (d), (e) and (f) we depict the non-interacting ($U = J = 0$) DOS that correspond to the systems (a) (b) and (c), respectively. Notice that increasing the bandwidth (even if just modestly by at most 15\%), paradoxically, leads to more robust ferromagnetism at \textit{lower U} --- a consequence of the van Hove singularity at the lower band edge. Also notice how when the next-nearest neighbor frustration is maximally reduced ($t_2 = 0$), the non-collinear 4-site magnetic order dominates the collinear one. Currently it is not clear to us why this would be the case, given how introducing $t_2, t_4$ seems to affect them equally based on pure lattice frustration arguments. }
	\label{fig:vary-t2m}
\end{figure*}

When it comes to  the hopping amplitudes $t_2$ and $t_4$,
given the spatial extent of the orbitals in the nickelates, it is
unrealistic to expect them to be larger than $t_1$. 
Hoppings bounded by $t_2, t_4 < 0.3$ only renormalize the overall
bandwidth in the non-interacting regime by at most 10\%, which should not be
enough to shift phase boundaries in the $U$-$J$ to any appreciable
degree. And yet even small adjustments to $t_2$ and $t_4$ can
significantly alter the magnetic phase diagram, as can be seen for
instance in Fig. \ref{fig:vary-t2m} for changing $t_2$ and in Fig.
\ref{fig:vary-t4m} for changing $t_4$. We believe these changes to be a
consequence of the shape of the DOS, which can change dramatically
even for small perturbations of the hopping parameters.
\cite{Peters2009b} The way the introduction of a new hopping
path affects the DOS can be predicted semi-analytically and depends,
in a hypercubic bipartite lattice, on whether the new hopping connects A-A
and B-B sublattice sites, or  the A-B sublattices.
\cite{Eckstein2005}

\begin{figure*}
	\includegraphics[width=0.85\textwidth]{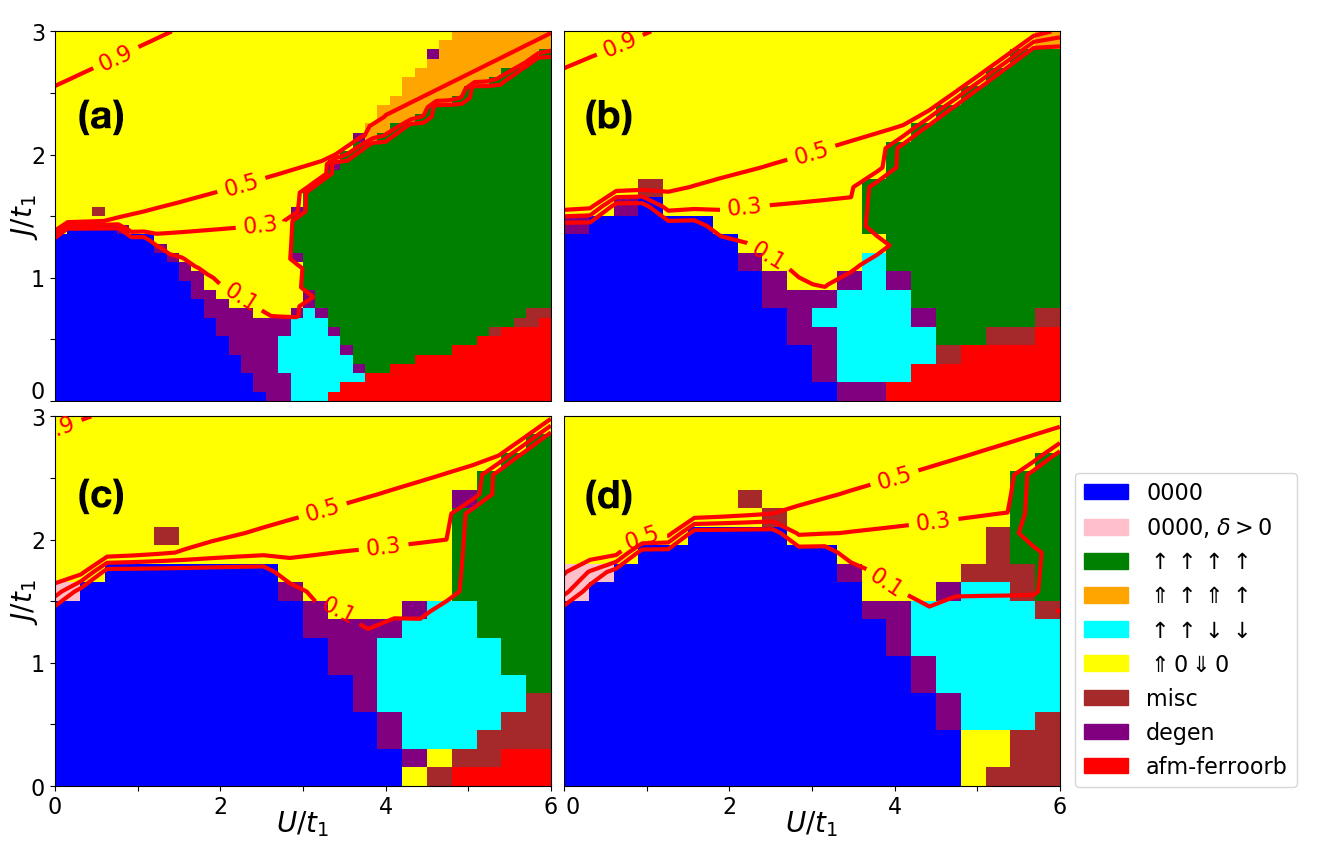}
	\caption{ Magnetic order in the HF ground state, as a function of $U$ and $J$, for $t_4 = 0, 0.1$ ((a) and (b), respectively), and $t_4 = 0.25, 0.35$ ((c) and (d), respectfully). The bandwidth is  $W = 6.4$ in all cases. Other parameters are $t_1 = 1, t_2 = 0.15, \epsilon_b = 0$. The growth of the metallic region is clearly not the effect of a renormalized bandwidth, but rather is due to the changes of the shape of the DOS. }
	\label{fig:vary-t4m}
\end{figure*}

In our case, $t_1$ and $t_4$ hopping appear to produce an entirely
symmetrical (about band centre) DOS: meanwhile, $t_2$ leads to a
strong asymmetry and the appearance of a van Hove singularity at
the lower band edge. In the spirit of the Stoner criterion, an
enhancement of the density of states near the Fermi level -- an
effective consequence of the introduction of a nonzero $t_2$ --
lowers the value of $U$ required for ferromagnetism to arise, thus
causing the metallic region near the origin to shrink significantly
(Fig. \ref{fig:vary-t2m}). Meanwhile, introducing a nonzero $t_4$ in
the presence of $t_2$ counteracts this tendency by boosting the
density of states above the Fermi level, thus increasing the
interaction strength $U$ required to exhibit magnetic order in the
ground state (see Fig. \ref{fig:vary-t4m}).

At the same time, the effect of $t_2$ and $t_4$ on the magnetic order
competition between the various 4-site contenders appears to be
negligible. Figure \ref{fig:energy-scalings-t2/t4} shows a typical
form of the energy modulation with $t_2$ and $t_4$ for the chief
contenders $\uparrow \uparrow \downarrow\downarrow$, $\uparrow\rightarrow\downarrow\leftarrow$ and $\Uparrow 0 \Downarrow 0$. The energies of the 4-site
states are affected in a very similar way, with none of them being the
clear favorite for the ground state. In the case of $t_4$, the
explanation for this is that the frustration costs introduced by the
4$^{\text{th}}$ neighbor-hopping, which connects sites that are two
lattice constants apart, are identical for all of these magnetic
states, so they are all equally disfavored. The effect for $t_2$ is
slightly different, as increasing it actually reduces the energy of
all of the magnetic states, initially showing a slight preference for
the non-collinear $\uparrow\rightarrow\downarrow\leftarrow$, then the collinear $\uparrow\uparrow\downarrow\downarrow$, and finally the
FM state. This can be understood in the spirit of the Stoner effect,
which is somewhat more general in this two-band Hubbard model: as the
density of states at the Fermi level grows, most kinds of magnetic
order benefit, but the FM state benefits the most. The strong response
of the FM state to the hopping amplitude modulations can be readily
seen in both Figs. \ref{fig:energy-scalings-t2/t4} (a) and (b): the
state suffers most strongly due to the added frustrations from $t_4$,
and benefits the most from the Stoner effect with $t_2$. Notice that
all of these effects occur with minimal bandwidth renormalization, as
discussed --- these are all purely consequences of the shape of the DOS.

Finally, we comment on the role played by the $\alpha$ anharmonicity parameter. Given
the form of the lattice energy $E_{\text{latt}} = \epsilon_b(
u_i^2 + \alpha u_i^4/2 )$, and a typical value for lattice distortion
$u_i \approx 0.5$, we see that the two terms compare numerically as
$\epsilon_b (0.25 + 0.03\alpha )$. Thus for $\alpha = 1$ the
quartic correction is merely 10\% of the quadratic contribution, and
serves only to modulate the bare magnitude of the lattice distortion
$u$, as determined by Eq. (\ref{hf5}), away from $u = \delta$, without
affecting the basic physics of the problem. The results shown in Fig.
\ref{fig:magnetism-a} confirm as much: insofar as the charge
modulation contours are displaced from their position at $\alpha = 0$,
it is the ones closest to $\delta=1$ and thus with the largest $u$ that are affected the most, whilst the
rest of the phase diagram remains the same. In the interest
of controlling the size of the lattice distortion, all the
calculations in this paper were carried out at $\alpha = 1$, unless
indicated otherwise.

\begin{figure}
	
	\includegraphics[width=0.48\textwidth]{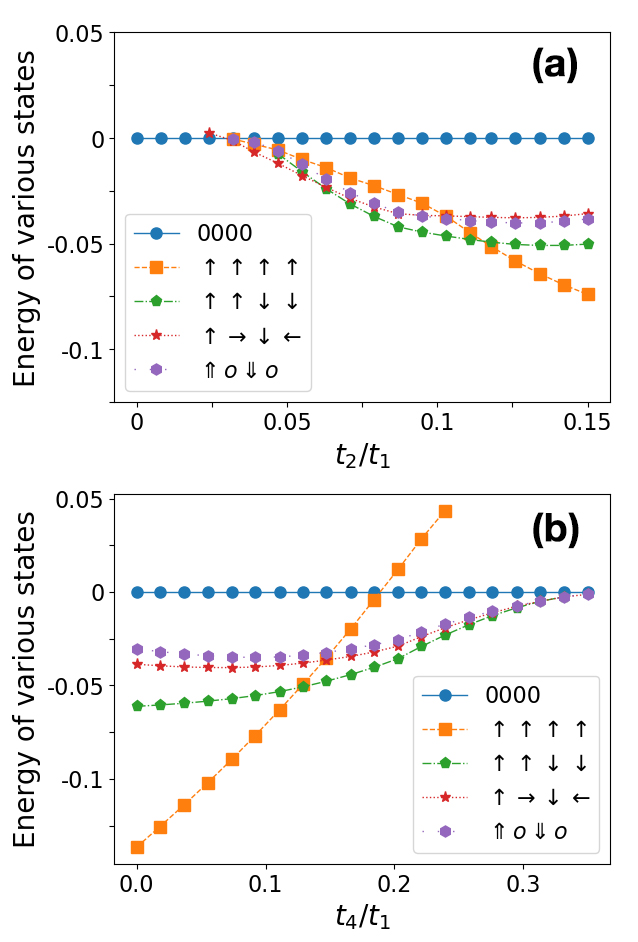}
	\caption{ Energies of several converged self-consistent HF
		states as a function of $t_4$, \textit{relative to} the energy of the metallic state. The different colors correspond to different magnetic orders (see the legend). The parameters are
		$U = 4, J = 0.316, t_1 = 1, t_4 = 0, \epsilon_b = 0$ for (a) and $U = 5, J = 0.316, t_1 = 1, t_2 = 0.15, \epsilon_b = 0$ for (b). Notice how in (b) the relative energies of the chief magnetic ground state contenders are not affected by the change in the hopping rate $t_4$ -- except for ferromagnetism, which gets strongly frustrated with the additional $t_4$ hopping and, paradoxically, ``unfrustrated'' with the introduction of $t_2$ hopping due to DOS effects (see the text for details). Meanwhile, in (a) with tuning the $t_2$ rate away from 0 the non-collinear $\uparrow\rightarrow\downarrow\leftarrow$ gets briefly favored, but then quickly loses out to the collinear $\uparrow\uparrow\downarrow\downarrow$, before ferromagnetism begins to reign supreme.}
	\label{fig:energy-scalings-t2/t4}
\end{figure}

\begin{figure}
	\includegraphics[width=0.48\textwidth]{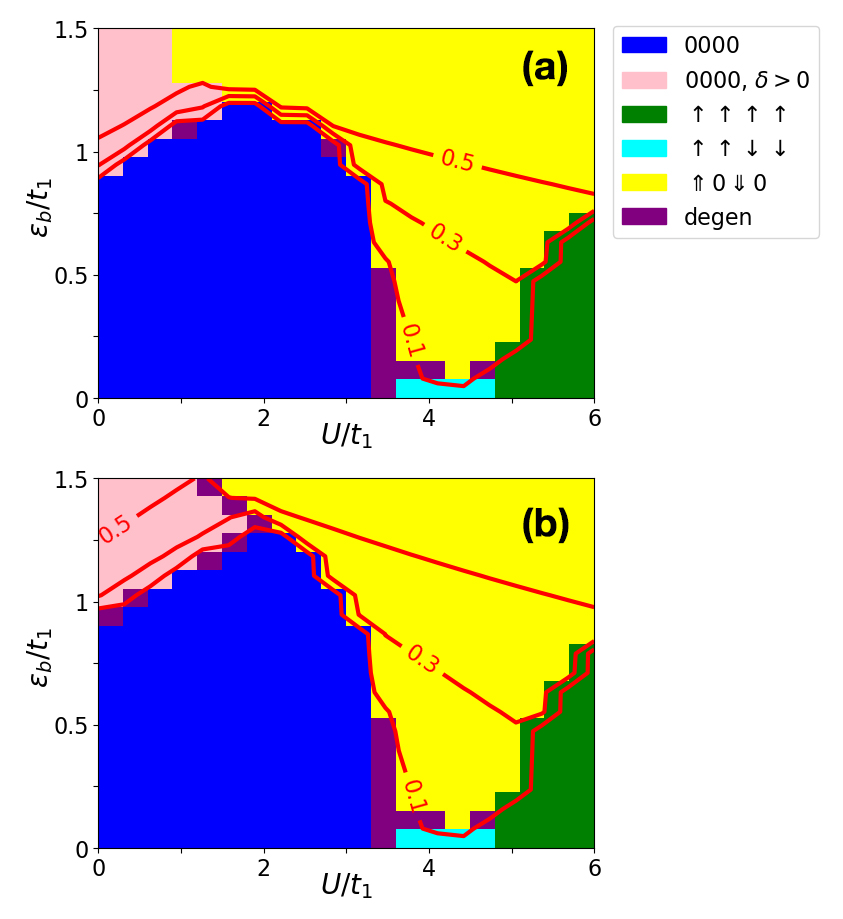}
	\caption{ Magnetic order in the HF ground state, as a function of $U$ and $\epsilon_b$, for $J/U=0.2$, and two different values of the anharmonicity $\alpha$: (a) $\alpha = 0$ and (b) $\alpha = 1$. Other parameters are $t_1 = 1, t_2 = 0.15, t_4 = 0.25$. }
	\label{fig:magnetism-a}
\end{figure}

\section{Conclusions}
 The magnetic order of the rare-earth nickelate series, much like the
 metal-insulator behavior and charge order, can be expected to couple
 to the lattice degrees of freedom. Even the simplest semiclassical
 version of the Holstein coupling is sufficient to aid charge
 disproportionation and to turn the material into an insulator in much of
 the parameter space. While several  magnetic orders
 are converge to self-consistency in our effective two-band Hubbard model
 for the nickelates, the $\Uparrow 0 \Downarrow 0$ antiferromagnetic order 
 dominates, usually presenting hand-in-hand with charge
 disproportionation $\delta\ne 0$. In contrast, the non-disproportionated
 collinear $\uparrow\uparrow\downarrow\downarrow$ and non-collinear $\uparrow\rightarrow\downarrow\leftarrow$ orders only arise at
 intermediate/large $U$ and small $J$ and do not fare well when the coupling
 to the lattice is increased, quickly disappearing from the phase diagram entirely.
 This can be easily understood, as the non-disproportionated modes
 cannot couple to the lattice distortion, given that for them $\delta = 0$
 and hence, via the self-consistency condition, also $u = 0$. Thus the
 main impact of the lattice on the magnetic order, in our model and at
 the HF level, is to make the $\Uparrow 0 \Downarrow 0$ order even more dominant, by
 decreasing its energy relative to that of the other states.
 Surprisingly, we find that self-consistency is never achieved for a
 state such as $\Uparrow\rightarrow\Downarrow\leftarrow$ or $\Uparrow\uparrow\Downarrow\downarrow$, that is, for $\delta \neq 0$.
 \cite{Scagnoli2006,Lu2018} Such states are unstable in the
 iterative sequences, with the slightest deviations from their
 expected mean-field parameter structure leading to fast flow towards
 the stable solution $\Uparrow 0 \Downarrow 0$.

While usually the charge modulated phase $\delta \neq 0$ always occurs with $\Uparrow
0 \Downarrow 0$ for small $U$, we found that introducing a finite
electron-lattice coupling $\epsilon_b$ also stabilizes a new phase, where the
charge modulation persisted on its own, without any associated
magnetic order. The magnetic order would then arise only
at higher $J$, leading to the effective decoupling of the charge
modulation and magnetic transitions -- a feature strongly reminiscent
of the canonical nickelates phase diagram.

We therefore conclude that all else being equal, coupling to the
lattice favors the $\Uparrow 0 \Downarrow 0$ order. However, one must keep in mind
that our simplified Hamiltonian may fail to capture properly some
aspects of the actual physics of these materials, especially in the
NCT scenario where the O sites should be included explicitly. Moreover, it is
known that the accuracy of the Hartree-Fock approximation can become
questionable as the strongly correlated limit is approached. One
caveat to note is that because of the negative charge transfer,
significant overlap between the (wide) O $2p$ bands and the Ni $3d$
bands should reduce the effective $U$ value (thus reducing the strength of correlations), explaining why many
in the literature have had success using a Hartree-Fock approach on
the nickelate problem.
\cite{Johnston2014,Garcia-Munoz1992b,Haule2017,Castellani1978} In
addition to the question of the validity of the mean-field approach,
other open questions remain. Specifically the nature of the orbitally
and magnetically ordered state we found at intermediate to large $U$ (\textit{i.e.} close to the Mott regime),
and other orbital orders possible within the model, are not entirely
clear and their investigation is left to future studies. Overall, strong $t_{pd}$ covalency places nickelates in the intermediate coupling regime of $U/W$ (for instance, see Ref. [\onlinecite{Mizokawa1995}]), wherein charge fluctuations coupled to lattice become an increasingly important factor in magnetic phase behavior of these materials.

\begin{acknowledgments}
The authors are grateful to George Sawatzky for valuable discussions about the model and the material series. We would also like to acknowledge the invaluable help of Evgenia Krichanovskaya with the graphic design of the various plots and diagrams. This work was supported by the UBC Stewart Blusson Quantum Matter Institute,  the Max-Planck-UBC-UTokyo Center for Quantum Materials and the Natural Sciences and Engineering Research Council of Canada. 
\end{acknowledgments}

\appendix

\section{Hopping operator}
The hopping operator is a sum of three terms
\begin{equation}
\hat{T} = \hat{T}_1 + \hat{T}_2 + \hat{T}_4,
\end{equation}
where
\begin{align}
\hat{T}_1 &= -t_1 \sum_{i\sigma} \sum_{\eta = x,y,z} \left( \hc{d}_{i\eta\sigma} d_{i+\eta,\eta\sigma} + \text{ h.c.}  \right), \\
\hat{T}_2 &= -t_2 \sum_{ \substack{i\eta\mu\sigma,\\ \eta \neq \mu \\ \eta,\mu=x,y,z} } \left(\hc{d}_{i+\mu+\eta,\mu\sigma} d_{i\eta\sigma} + \text{h.c.} \right), \\
\hat{T}_4 &= -t_4 \sum_{i\sigma} \sum_{\eta = x,y,z} \left( \hc{d}_{i\eta\sigma} d_{i+2\eta,\eta\sigma} + \text{ h.c.}  \right).
\end{align}

\begin{widetext}
After a Fourier transform and factorization, the hopping operator has the form
\begin{equation}
\hat{T} = \sum_{\mathbf{k}\sigma} \Big[ t_{zz}(\mathbf{k}) \hc{d}_{\mathbf{k}z\sigma} d_{\mathbf{k}z\sigma} + t_{\bar{z}\bar{z}}(\mathbf{k}) \hc{d}_{\mathbf{k}\bar{z}\sigma} d_{\mathbf{k}\bar{z}\sigma}  + t_{z\bar{z}}(\mathbf{k}) (\hc{d}_{\mathbf{k}z\sigma} d_{\mathbf{k}\bar{z}\sigma} + \hc{d}_{\mathbf{k}\bar{z}\sigma} d_{\mathbf{k}z\sigma} ) \Big],
\end{equation}
with the definitions
\begin{align}\nonumber
t_{zz}(\mathbf{k}) = &-2 t_1\left( \cos(k_z) + \frac{1}{4} [\cos(k_x)  + \cos(k_y) ] \right)- 2t_4 \left( \cos(2k_z) + \frac{1}{4} [\cos(2k_x)  + \cos(2k_y) ] \right) - \\
&-2t_2 \left( \cos(k_x) \cos(k_y) - 2\cos(k_z)(\cos(k_y) + \cos(k_x)) \right), \nonumber\\
t_{\bar{z}\bar{z}}(\mathbf{k}) = &-\frac{3t_1}{2} [\cos(k_x)  + \cos(k_y) ] -\frac{3t_4}{2} [\cos(2k_x)  + \cos(2k_y) ] + 6 t_2\cos(k_x) \cos(k_y),\nonumber
\\
t_{z\bar{z}}(\mathbf{k}) = &\frac{\sqrt{3}t_1}{2} [\cos(k_x)  - \cos(k_y) ] + \frac{\sqrt{3}t_4}{2} [\cos(2k_x)  - \cos(2k_y) ] -2\sqrt{3}t_2  \cos(k_z)( \cos(k_x) - \cos(k_y)) .\nonumber
\end{align}
\end{widetext}

\section{Minimizing lattice contributions}
\label{app:latt-self-consis}
In the text we were faced with the need to solve the cubic equation
\begin{align}
u^3 + \frac{1}{\alpha}u - \frac{\delta}{\alpha} = 0
\end{align}
which arose during lattice energy minimization in the Hartree-Fock process. As we mentioned, it admits a closed form solution using Cardano's formula. More explicitly, write
\[
u^3 = (s - t)^3, \quad \frac{1}{\alpha} = 3st, \quad \frac{\delta}{\alpha} = s^3 - t^3
\]
Combining the latter two equations into one for $t$, we find
\[
t^6 + t^3\left( \frac{\delta}{\alpha} \right) - \left( \frac{1}{3\alpha} \right)^3 = 0
\]
which is easily solved as a quadratic
\[
t^3 = \frac{1}{2}\left(-\frac{\delta}{\alpha} \pm \sqrt{\left( \frac{\delta}{\alpha} \right)^2 + \frac{4}{27\alpha^3}}\right)
\]
\[
s^3 = t^3 + \frac{\delta}{\alpha} = \frac{1}{2}\left(\frac{\delta}{\alpha} \pm \sqrt{\left( \frac{\delta}{\alpha} \right)^2 + \frac{4}{27\alpha^3}}\right)
\]
from which we can recover the expression for $u$ in Eq. (\ref{hf5}).

\section{Full mean-field Hamiltonian}
The full effective Hamiltonian is:
\begin{widetext}
\begin{multline}
\hat{H}_{\text{HF}} = \sum_{\mathbf{k}ab\sigma} t_{ab}(\mathbf{k}) \hc{d}_{\mathbf{k}a\sigma} d_{\mathbf{k}b\sigma} + \sum_{\mathbf{k}a\sigma} \Big( \frac{3U-5J}{4} - \frac{\sigma}{2} (U+J) S_{\text{FM}} \Big) \hc{d}_{\mathbf{k}a\sigma} d_{\mathbf{k}a\sigma} + \nonumber\\
+ \Big( \frac{3U-5J}{4}\delta - 2\epsilon_b u - \frac{\sigma}{2} (U+J) S_{\text{AFM}} \Big) \hc{d}_{\mathbf{k}+\mathbf{Q}_c,a\sigma} d_{\mathbf{k}a\sigma} - \nonumber\\
-\frac{\sigma}{4}(U+J) \Big( (S_{\text{1z}} - i S_{\text{2z}}) \hc{d}_{\mathbf{k}+\mathbf{Q}_m,a\sigma} d_{\mathbf{k}a\sigma} + (S_{\text{1z}} + iS_{\text{2z}}) \hc{d}_{\mathbf{k}-\mathbf{Q}_m,a\sigma} d_{\mathbf{k}a\sigma}  \Big) - \nonumber\\
- \frac{U+J}{4} \Big( (S_{\text{1x}} - i S_{\text{2x}}) \hc{d}_{\mathbf{k}+\mathbf{Q}_m,a\sigma} d_{\mathbf{k}a\bar{\sigma}} + (S_{\text{1x}} + iS_{\text{2x}}) \hc{d}_{\mathbf{k}-\mathbf{Q}_m,a\sigma} d_{\mathbf{k}a\bar{\sigma}}  \Big) \nonumber\\
- \frac{U-J}{2} \Big( (X_{\text{1}} - i X_{\text{2}}) \hc{d}_{\mathbf{k}+\mathbf{Q}_m,a\sigma} d_{\mathbf{k}\bar{a}\bar{\sigma}} + (S_{\text{1x}} + iS_{\text{2x}}) \hc{d}_{\mathbf{k}-\mathbf{Q}_m,a\sigma} d_{\mathbf{k}\bar{a}\bar{\sigma}}  \Big) + \nonumber\\
+ \left[ (5J-U) O_1 - \sigma(U-J)Z_1\right] \hc{d}_{\mathbf{k}a\sigma} d_{\mathbf{k}\bar{a}\sigma} + \left[ (5J-U) O_2 - \sigma(U-J)Z_2\right] \hc{d}_{\mathbf{k}+\mathbf{Q}_c,a\sigma} d_{\mathbf{k}\bar{a}\sigma} \nonumber\\
- \frac{\sigma}{2} (U-J) \left[ (Z_3-iZ_4) \hc{d}_{\mathbf{k}+\mathbf{Q}_m,a\sigma} d_{\mathbf{k}\bar{a}\sigma} + (Z_3 + iZ_4) \hc{d}_{\mathbf{k}-\mathbf{Q}_m,a\sigma} d_{\mathbf{k}\bar{a}\sigma}  \right].
\label{eq:HF-ham}
\end{multline}

\section{Hartree-Fock energy}
The full expression for the electron-electron interactions part of the Hamiltonian in a Hartree-Fock state $\ket{\Psi_e}$ is given by
\begin{multline}
\frac{1}{N} \langle \hat{H}_e \rangle = \frac{3U-5J}{8} (1+\delta^2) 
- \frac{U+J}{2} \Big( S_{\text{FM}}^2  + S_{\text{AFM}}^2 + \frac{S_{\text{1x}}^2 + S_{\text{1z}}^2 + S_{\text{2x}}^2 + S_{\text{2z}}^2}{2} \Big) 
- 2 (U - 5J) (O_1^2 + O_2^2) - \\
 - 2 (U - J) \left(  Z_1^2 + Z_2^2 + \frac{Z_3^2 + Z_4^2}{2} \right) 
 - (U - J) (X_1^2 + X_2^2).
\end{multline}
\end{widetext}

\bibliographystyle{apsrev4-1}

\begin{thebibliography}{40}%
	\makeatletter
	\providecommand \@ifxundefined [1]{%
		\@ifx{#1\undefined}
	}%
	\providecommand \@ifnum [1]{%
		\ifnum #1\expandafter \@firstoftwo
		\else \expandafter \@secondoftwo
		\fi
	}%
	\providecommand \@ifx [1]{%
		\ifx #1\expandafter \@firstoftwo
		\else \expandafter \@secondoftwo
		\fi
	}%
	\providecommand \natexlab [1]{#1}%
	\providecommand \enquote  [1]{``#1''}%
	\providecommand \bibnamefont  [1]{#1}%
	\providecommand \bibfnamefont [1]{#1}%
	\providecommand \citenamefont [1]{#1}%
	\providecommand \href@noop [0]{\@secondoftwo}%
	\providecommand \href [0]{\begingroup \@sanitize@url \@href}%
	\providecommand \@href[1]{\@@startlink{#1}\@@href}%
	\providecommand \@@href[1]{\endgroup#1\@@endlink}%
	\providecommand \@sanitize@url [0]{\catcode `\\12\catcode `\$12\catcode
		`\&12\catcode `\#12\catcode `\^12\catcode `\_12\catcode `\%12\relax}%
	\providecommand \@@startlink[1]{}%
	\providecommand \@@endlink[0]{}%
	\providecommand \url  [0]{\begingroup\@sanitize@url \@url }%
	\providecommand \@url [1]{\endgroup\@href {#1}{\urlprefix }}%
	\providecommand \urlprefix  [0]{URL }%
	\providecommand \Eprint [0]{\href }%
	\providecommand \doibase [0]{http://dx.doi.org/}%
	\providecommand \selectlanguage [0]{\@gobble}%
	\providecommand \bibinfo  [0]{\@secondoftwo}%
	\providecommand \bibfield  [0]{\@secondoftwo}%
	\providecommand \translation [1]{[#1]}%
	\providecommand \BibitemOpen [0]{}%
	\providecommand \bibitemStop [0]{}%
	\providecommand \bibitemNoStop [0]{.\EOS\space}%
	\providecommand \EOS [0]{\spacefactor3000\relax}%
	\providecommand \BibitemShut  [1]{\csname bibitem#1\endcsname}%
	\let\auto@bib@innerbib\@empty
	\bibitem [{\citenamefont {Catalan}(2008)}]{Catalan2008}%
	\BibitemOpen
	\bibfield  {author} {\bibinfo {author} {\bibfnamefont {G.}~\bibnamefont
			{Catalan}},\ }\href {\doibase 10.1080/01411590801992463} {\bibfield
		{journal} {\bibinfo  {journal} {Phase Transitions}\ }\textbf {\bibinfo
			{volume} {81}},\ \bibinfo {pages} {729} (\bibinfo {year} {2008})}\BibitemShut
	{NoStop}%
	\bibitem [{\citenamefont {Medarde}(1997)}]{Medarde1997}%
	\BibitemOpen
	\bibfield  {author} {\bibinfo {author} {\bibfnamefont {M.~L.}\ \bibnamefont
			{Medarde}},\ }\href {\doibase 10.1088/0953-8984/9/8/003} {\bibfield
		{journal} {\bibinfo  {journal} {Journal of Physics: Condensed Matter}\
		}\textbf {\bibinfo {volume} {9}},\ \bibinfo {pages} {1679} (\bibinfo {year}
		{1997})}\BibitemShut {NoStop}%
	\bibitem [{\citenamefont {Hepting}(2017)}]{Hepting2017}%
	\BibitemOpen
	\bibfield  {author} {\bibinfo {author} {\bibfnamefont {M.}~\bibnamefont
			{Hepting}},\ }\href {\doibase 10.1007/978-3-319-60531-9} {\emph {\bibinfo
			{title} {{Ordering Phenomena in Rare-Earth Nickelate Heterostructures}}}},\
	Springer Theses\ (\bibinfo  {publisher} {Springer International Publishing},\
	\bibinfo {address} {Cham},\ \bibinfo {year} {2017})\BibitemShut {NoStop}%
	\bibitem [{\citenamefont {Giovannetti}\ \emph {et~al.}(2009)\citenamefont
		{Giovannetti}, \citenamefont {Kumar}, \citenamefont {Khomskii}, \citenamefont
		{Picozzi},\ and\ \citenamefont {van~den Brink}}]{Giovannetti2009}%
	\BibitemOpen
	\bibfield  {author} {\bibinfo {author} {\bibfnamefont {G.}~\bibnamefont
			{Giovannetti}}, \bibinfo {author} {\bibfnamefont {S.}~\bibnamefont {Kumar}},
		\bibinfo {author} {\bibfnamefont {D.}~\bibnamefont {Khomskii}}, \bibinfo
		{author} {\bibfnamefont {S.}~\bibnamefont {Picozzi}}, \ and\ \bibinfo
		{author} {\bibfnamefont {J.}~\bibnamefont {van~den Brink}},\ }\href {\doibase
		10.1103/PhysRevLett.103.156401} {\bibfield  {journal} {\bibinfo  {journal}
			{Physical Review Letters}\ }\textbf {\bibinfo {volume} {103}},\ \bibinfo
		{pages} {156401} (\bibinfo {year} {2009})}\BibitemShut {NoStop}%
	\bibitem [{\citenamefont {Spaldin}(2017)}]{Spaldin2017}%
	\BibitemOpen
	\bibfield  {author} {\bibinfo {author} {\bibfnamefont {N.~A.}\ \bibnamefont
			{Spaldin}},\ }\href {\doibase 10.1557/mrs.2017.86} {\bibfield  {journal}
		{\bibinfo  {journal} {MRS Bulletin}\ }\textbf {\bibinfo {volume} {42}},\
		\bibinfo {pages} {385} (\bibinfo {year} {2017})}\BibitemShut {NoStop}%
	\bibitem [{\citenamefont {Scott}(2007)}]{Scott2007}%
	\BibitemOpen
	\bibfield  {author} {\bibinfo {author} {\bibfnamefont {J.~F.}\ \bibnamefont
			{Scott}},\ }\href {\doibase 10.1038/nmat1868} {\bibfield  {journal} {\bibinfo
			{journal} {Nature Materials}\ }\textbf {\bibinfo {volume} {6}},\ \bibinfo
		{pages} {256} (\bibinfo {year} {2007})}\BibitemShut {NoStop}%
	\bibitem [{Note1()}]{Note1}%
	\BibitemOpen
	\bibinfo {note} {These octahedral cages undergo a variety of tilts, twists
		are other complicated distortions due to the rare-earth ion being too small
		to accommodate a true perovskite lattice. However, such distortions change
		little across the temperature range we are interested in: the main change to
		lattice structure comes from the breathing mode distortion. As such, for the
		purposes of our analysis we can focus on the breathing-mode
		distortion.}\BibitemShut {Stop}%
	\bibitem [{\citenamefont {Rodr{\'{i}}guez-Carvajal}\ \emph
		{et~al.}(1998)\citenamefont {Rodr{\'{i}}guez-Carvajal}, \citenamefont
		{Rosenkranz}, \citenamefont {Medarde}, \citenamefont {Lacorre}, \citenamefont
		{Fernandez-D{\'{i}}az}, \citenamefont {Fauth},\ and\ \citenamefont
		{Trounov}}]{Rodriguez-Carvajal1998}%
	\BibitemOpen
	\bibfield  {author} {\bibinfo {author} {\bibfnamefont {J.}~\bibnamefont
			{Rodr{\'{i}}guez-Carvajal}}, \bibinfo {author} {\bibfnamefont
			{S.}~\bibnamefont {Rosenkranz}}, \bibinfo {author} {\bibfnamefont
			{M.}~\bibnamefont {Medarde}}, \bibinfo {author} {\bibfnamefont
			{P.}~\bibnamefont {Lacorre}}, \bibinfo {author} {\bibfnamefont {M.~T.}\
			\bibnamefont {Fernandez-D{\'{i}}az}}, \bibinfo {author} {\bibfnamefont
			{F.}~\bibnamefont {Fauth}}, \ and\ \bibinfo {author} {\bibfnamefont
			{V.}~\bibnamefont {Trounov}},\ }\href {\doibase 10.1103/PhysRevB.57.456}
	{\bibfield  {journal} {\bibinfo  {journal} {Physical Review B}\ }\textbf
		{\bibinfo {volume} {57}},\ \bibinfo {pages} {456} (\bibinfo {year}
		{1998})}\BibitemShut {NoStop}%
	\bibitem [{\citenamefont {Scagnoli}\ \emph {et~al.}(2006)\citenamefont
		{Scagnoli}, \citenamefont {Staub}, \citenamefont {Mulders}, \citenamefont
		{Janousch}, \citenamefont {Meijer}, \citenamefont {Hammerl}, \citenamefont
		{Tonnerre},\ and\ \citenamefont {Stojic}}]{Scagnoli2006}%
	\BibitemOpen
	\bibfield  {author} {\bibinfo {author} {\bibfnamefont {V.}~\bibnamefont
			{Scagnoli}}, \bibinfo {author} {\bibfnamefont {U.}~\bibnamefont {Staub}},
		\bibinfo {author} {\bibfnamefont {A.~M.}\ \bibnamefont {Mulders}}, \bibinfo
		{author} {\bibfnamefont {M.}~\bibnamefont {Janousch}}, \bibinfo {author}
		{\bibfnamefont {G.~I.}\ \bibnamefont {Meijer}}, \bibinfo {author}
		{\bibfnamefont {G.}~\bibnamefont {Hammerl}}, \bibinfo {author} {\bibfnamefont
			{J.~M.}\ \bibnamefont {Tonnerre}}, \ and\ \bibinfo {author} {\bibfnamefont
			{N.}~\bibnamefont {Stojic}},\ }\href {\doibase 10.1103/PhysRevB.73.100409}
	{\bibfield  {journal} {\bibinfo  {journal} {Physical Review B}\ }\textbf
		{\bibinfo {volume} {73}},\ \bibinfo {pages} {100409(R)} (\bibinfo {year}
		{2006})}\BibitemShut {NoStop}%
	\bibitem [{\citenamefont {Medarde}\ \emph {et~al.}(2009)\citenamefont
		{Medarde}, \citenamefont {Dallera}, \citenamefont {Grioni}, \citenamefont
		{Delley}, \citenamefont {Vernay}, \citenamefont {Mesot}, \citenamefont
		{Sikora}, \citenamefont {Alonso},\ and\ \citenamefont
		{Mart{\'{i}}nez-Lope}}]{Medarde2009}%
	\BibitemOpen
	\bibfield  {author} {\bibinfo {author} {\bibfnamefont {M.}~\bibnamefont
			{Medarde}}, \bibinfo {author} {\bibfnamefont {C.}~\bibnamefont {Dallera}},
		\bibinfo {author} {\bibfnamefont {M.}~\bibnamefont {Grioni}}, \bibinfo
		{author} {\bibfnamefont {B.}~\bibnamefont {Delley}}, \bibinfo {author}
		{\bibfnamefont {F.}~\bibnamefont {Vernay}}, \bibinfo {author} {\bibfnamefont
			{J.}~\bibnamefont {Mesot}}, \bibinfo {author} {\bibfnamefont
			{M.}~\bibnamefont {Sikora}}, \bibinfo {author} {\bibfnamefont {J.~A.}\
			\bibnamefont {Alonso}}, \ and\ \bibinfo {author} {\bibfnamefont {M.~J.}\
			\bibnamefont {Mart{\'{i}}nez-Lope}},\ }\href {\doibase
		10.1103/PhysRevB.80.245105} {\bibfield  {journal} {\bibinfo  {journal}
			{Physical Review B}\ }\textbf {\bibinfo {volume} {80}},\ \bibinfo {pages}
		{245105} (\bibinfo {year} {2009})}\BibitemShut {NoStop}%
	\bibitem [{\citenamefont {Mizokawa}\ \emph {et~al.}(2000)\citenamefont
		{Mizokawa}, \citenamefont {Khomskii},\ and\ \citenamefont
		{Sawatzky}}]{Mizokawa2000}%
	\BibitemOpen
	\bibfield  {author} {\bibinfo {author} {\bibfnamefont {T.}~\bibnamefont
			{Mizokawa}}, \bibinfo {author} {\bibfnamefont {D.~I.}\ \bibnamefont
			{Khomskii}}, \ and\ \bibinfo {author} {\bibfnamefont {G.~A.}\ \bibnamefont
			{Sawatzky}},\ }\href {\doibase 10.1103/PhysRevB.61.11263} {\bibfield
		{journal} {\bibinfo  {journal} {Physical Review B}\ }\textbf {\bibinfo
			{volume} {61}},\ \bibinfo {pages} {11263} (\bibinfo {year}
		{2000})}\BibitemShut {NoStop}%
	\bibitem [{\citenamefont {Park}\ \emph {et~al.}(2012)\citenamefont {Park},
		\citenamefont {Millis},\ and\ \citenamefont {Marianetti}}]{Park2012}%
	\BibitemOpen
	\bibfield  {author} {\bibinfo {author} {\bibfnamefont {H.}~\bibnamefont
			{Park}}, \bibinfo {author} {\bibfnamefont {A.~J.}\ \bibnamefont {Millis}}, \
		and\ \bibinfo {author} {\bibfnamefont {C.~A.}\ \bibnamefont {Marianetti}},\
	}\href {\doibase 10.1103/PhysRevLett.109.156402} {\bibfield  {journal}
		{\bibinfo  {journal} {Physical Review Letters}\ }\textbf {\bibinfo {volume}
			{109}},\ \bibinfo {pages} {156402} (\bibinfo {year} {2012})}\BibitemShut
	{NoStop}%
	\bibitem [{\citenamefont {Lau}\ and\ \citenamefont {Millis}(2013)}]{Lau2013}%
	\BibitemOpen
	\bibfield  {author} {\bibinfo {author} {\bibfnamefont {B.}~\bibnamefont
			{Lau}}\ and\ \bibinfo {author} {\bibfnamefont {A.~J.}\ \bibnamefont
			{Millis}},\ }\href {\doibase 10.1103/PhysRevLett.110.126404} {\bibfield
		{journal} {\bibinfo  {journal} {Physical Review Letters}\ }\textbf {\bibinfo
			{volume} {110}},\ \bibinfo {pages} {126404} (\bibinfo {year}
		{2013})}\BibitemShut {NoStop}%
	\bibitem [{\citenamefont {Puggioni}\ \emph {et~al.}(2012)\citenamefont
		{Puggioni}, \citenamefont {Filippetti},\ and\ \citenamefont
		{Fiorentini}}]{Puggioni2012}%
	\BibitemOpen
	\bibfield  {author} {\bibinfo {author} {\bibfnamefont {D.}~\bibnamefont
			{Puggioni}}, \bibinfo {author} {\bibfnamefont {A.}~\bibnamefont
			{Filippetti}}, \ and\ \bibinfo {author} {\bibfnamefont {V.}~\bibnamefont
			{Fiorentini}},\ }\href {\doibase 10.1103/PhysRevB.86.195132} {\bibfield
		{journal} {\bibinfo  {journal} {Physical Review B}\ }\textbf {\bibinfo
			{volume} {86}},\ \bibinfo {pages} {195132} (\bibinfo {year}
		{2012})}\BibitemShut {NoStop}%
	\bibitem [{\citenamefont {Caviglia}\ \emph {et~al.}(2012)\citenamefont
		{Caviglia}, \citenamefont {Scherwitzl}, \citenamefont {Popovich},
		\citenamefont {Hu}, \citenamefont {Bromberger}, \citenamefont {Singla},
		\citenamefont {Mitrano}, \citenamefont {Hoffmann}, \citenamefont {Kaiser},
		\citenamefont {Zubko}, \citenamefont {Gariglio}, \citenamefont {Triscone},
		\citenamefont {F{\"{o}}rst},\ and\ \citenamefont {Cavalleri}}]{Caviglia2012}%
	\BibitemOpen
	\bibfield  {author} {\bibinfo {author} {\bibfnamefont {A.~D.}\ \bibnamefont
			{Caviglia}}, \bibinfo {author} {\bibfnamefont {R.}~\bibnamefont
			{Scherwitzl}}, \bibinfo {author} {\bibfnamefont {P.}~\bibnamefont
			{Popovich}}, \bibinfo {author} {\bibfnamefont {W.}~\bibnamefont {Hu}},
		\bibinfo {author} {\bibfnamefont {H.}~\bibnamefont {Bromberger}}, \bibinfo
		{author} {\bibfnamefont {R.}~\bibnamefont {Singla}}, \bibinfo {author}
		{\bibfnamefont {M.}~\bibnamefont {Mitrano}}, \bibinfo {author} {\bibfnamefont
			{M.~C.}\ \bibnamefont {Hoffmann}}, \bibinfo {author} {\bibfnamefont
			{S.}~\bibnamefont {Kaiser}}, \bibinfo {author} {\bibfnamefont
			{P.}~\bibnamefont {Zubko}}, \bibinfo {author} {\bibfnamefont
			{S.}~\bibnamefont {Gariglio}}, \bibinfo {author} {\bibfnamefont {J.-M.}\
			\bibnamefont {Triscone}}, \bibinfo {author} {\bibfnamefont {M.}~\bibnamefont
			{F{\"{o}}rst}}, \ and\ \bibinfo {author} {\bibfnamefont {A.}~\bibnamefont
			{Cavalleri}},\ }\href {\doibase 10.1103/PhysRevLett.108.136801} {\bibfield
		{journal} {\bibinfo  {journal} {Physical Review Letters}\ }\textbf {\bibinfo
			{volume} {108}},\ \bibinfo {pages} {136801} (\bibinfo {year}
		{2012})}\BibitemShut {NoStop}%
	\bibitem [{\citenamefont {Garc{\'{i}}a-Mu{\~{n}}oz}\ \emph
		{et~al.}(1992{\natexlab{a}})\citenamefont {Garc{\'{i}}a-Mu{\~{n}}oz},
		\citenamefont {Rodr{\'{i}}guez-Carvajal}, \citenamefont {Lacorre},\ and\
		\citenamefont {Torrance}}]{Garcia-Munoz1992a}%
	\BibitemOpen
	\bibfield  {author} {\bibinfo {author} {\bibfnamefont {J.~L.}\ \bibnamefont
			{Garc{\'{i}}a-Mu{\~{n}}oz}}, \bibinfo {author} {\bibfnamefont
			{J.}~\bibnamefont {Rodr{\'{i}}guez-Carvajal}}, \bibinfo {author}
		{\bibfnamefont {P.}~\bibnamefont {Lacorre}}, \ and\ \bibinfo {author}
		{\bibfnamefont {J.~B.}\ \bibnamefont {Torrance}},\ }\href {\doibase
		10.1103/PhysRevB.46.4414} {\bibfield  {journal} {\bibinfo  {journal}
			{Physical Review B}\ }\textbf {\bibinfo {volume} {46}},\ \bibinfo {pages}
		{4414} (\bibinfo {year} {1992}{\natexlab{a}})}\BibitemShut {NoStop}%
	\bibitem [{\citenamefont {Garc{\'{i}}a-Mu{\~{n}}oz}\ \emph
		{et~al.}(1992{\natexlab{b}})\citenamefont {Garc{\'{i}}a-Mu{\~{n}}oz},
		\citenamefont {Rodr{\'{i}}guez-Carvajal},\ and\ \citenamefont
		{Lacorre}}]{Garcia-Munoz1992b}%
	\BibitemOpen
	\bibfield  {author} {\bibinfo {author} {\bibfnamefont {J.~L.}\ \bibnamefont
			{Garc{\'{i}}a-Mu{\~{n}}oz}}, \bibinfo {author} {\bibfnamefont
			{J.}~\bibnamefont {Rodr{\'{i}}guez-Carvajal}}, \ and\ \bibinfo {author}
		{\bibfnamefont {P.}~\bibnamefont {Lacorre}},\ }\href {\doibase
		10.1209/0295-5075/20/3/009} {\bibfield  {journal} {\bibinfo  {journal}
			{Europhysics Letters (EPL)}\ }\textbf {\bibinfo {volume} {20}},\ \bibinfo
		{pages} {241} (\bibinfo {year} {1992}{\natexlab{b}})}\BibitemShut {NoStop}%
	\bibitem [{\citenamefont {Mu{\~{n}}oz}\ \emph {et~al.}(2009)\citenamefont
		{Mu{\~{n}}oz}, \citenamefont {Alonso}, \citenamefont {Mart{\'{i}}nez-Lope},\
		and\ \citenamefont {Fern{\'{a}}ndez-D{\'{i}}az}}]{Munoz2009}%
	\BibitemOpen
	\bibfield  {author} {\bibinfo {author} {\bibfnamefont {A.}~\bibnamefont
			{Mu{\~{n}}oz}}, \bibinfo {author} {\bibfnamefont {J.}~\bibnamefont {Alonso}},
		\bibinfo {author} {\bibfnamefont {M.}~\bibnamefont {Mart{\'{i}}nez-Lope}}, \
		and\ \bibinfo {author} {\bibfnamefont {M.}~\bibnamefont
			{Fern{\'{a}}ndez-D{\'{i}}az}},\ }\href {\doibase 10.1016/j.jssc.2009.05.013}
	{\bibfield  {journal} {\bibinfo  {journal} {Journal of Solid State
				Chemistry}\ }\textbf {\bibinfo {volume} {182}},\ \bibinfo {pages} {1982}
		(\bibinfo {year} {2009})}\BibitemShut {NoStop}%
	\bibitem [{\citenamefont {Haule}\ and\ \citenamefont
		{Pascut}(2017)}]{Haule2017}%
	\BibitemOpen
	\bibfield  {author} {\bibinfo {author} {\bibfnamefont {K.}~\bibnamefont
			{Haule}}\ and\ \bibinfo {author} {\bibfnamefont {G.~L.}\ \bibnamefont
			{Pascut}},\ }\href {\doibase 10.1038/s41598-017-10374-2} {\bibfield
		{journal} {\bibinfo  {journal} {Scientific Reports}\ }\textbf {\bibinfo
			{volume} {7}},\ \bibinfo {pages} {10375} (\bibinfo {year}
		{2017})}\BibitemShut {NoStop}%
	\bibitem [{\citenamefont {Lee}\ \emph {et~al.}(2011)\citenamefont {Lee},
		\citenamefont {Chen},\ and\ \citenamefont {Balents}}]{Lee2011}%
	\BibitemOpen
	\bibfield  {author} {\bibinfo {author} {\bibfnamefont {S.~B.}~\bibnamefont
			{Lee}}, \bibinfo {author} {\bibfnamefont {R.}~\bibnamefont {Chen}}, \ and\
		\bibinfo {author} {\bibfnamefont {L.}~\bibnamefont {Balents}},\ }\href
	{https://journals.aps.org/prb/pdf/10.1103/PhysRevB.84.165119} {\bibfield
		{journal} {\bibinfo  {journal} {Physical Review B}\ }\textbf {\bibinfo
			{volume} {84}},\ \bibinfo {pages} {165119} (\bibinfo {year} {2011})}\BibitemShut {NoStop}%
	\bibitem [{\citenamefont {Lu}\ \emph {et~al.}(2018)\citenamefont {Lu},
		\citenamefont {Betto}, \citenamefont {F{\"{u}}rsich}, \citenamefont {Suzuki},
		\citenamefont {Kim}, \citenamefont {Cristiani}, \citenamefont {Logvenov},
		\citenamefont {Brookes}, \citenamefont {Benckiser}, \citenamefont
		{Haverkort}, \citenamefont {Khaliullin}, \citenamefont {{Le Tacon}},
		\citenamefont {Minola},\ and\ \citenamefont {Keimer}}]{Lu2018}%
	\BibitemOpen
	\bibfield  {author} {\bibinfo {author} {\bibfnamefont {Y.}~\bibnamefont
			{Lu}}, \bibinfo {author} {\bibfnamefont {D.}~\bibnamefont {Betto}}, \bibinfo
		{author} {\bibfnamefont {K.}~\bibnamefont {F{\"{u}}rsich}}, \bibinfo {author}
		{\bibfnamefont {H.}~\bibnamefont {Suzuki}}, \bibinfo {author} {\bibfnamefont
			{H.-H.}\ \bibnamefont {Kim}}, \bibinfo {author} {\bibfnamefont
			{G.}~\bibnamefont {Cristiani}}, \bibinfo {author} {\bibfnamefont
			{G.}~\bibnamefont {Logvenov}}, \bibinfo {author} {\bibfnamefont {N.~B.}\
			\bibnamefont {Brookes}}, \bibinfo {author} {\bibfnamefont {E.}~\bibnamefont
			{Benckiser}}, \bibinfo {author} {\bibfnamefont {M.~W.}\ \bibnamefont
			{Haverkort}}, \bibinfo {author} {\bibfnamefont {G.}~\bibnamefont
			{Khaliullin}}, \bibinfo {author} {\bibfnamefont {M.}~\bibnamefont {{Le
					Tacon}}}, \bibinfo {author} {\bibfnamefont {M.}~\bibnamefont {Minola}}, \
		and\ \bibinfo {author} {\bibfnamefont {B.}~\bibnamefont {Keimer}},\ }\href
	{https://journals.aps.org/prx/pdf/10.1103/PhysRevX.8.031014} {\bibfield
		{journal} {\bibinfo  {journal} {Physical Review X}\ }\textbf {\bibinfo
			{volume} {8}},\ \bibinfo {pages} {031014} (\bibinfo {year} {2018})}\BibitemShut {NoStop}%
	\bibitem [{\citenamefont {Kanamori}(1963)}]{Kanamori1963}%
	\BibitemOpen
	\bibfield  {author} {\bibinfo {author} {\bibfnamefont {J.}~\bibnamefont
			{Kanamori}},\ }\href {https://doi.org/10.1143/PTP.30.275} {\bibfield
		{journal} {\bibinfo  {journal} {Progress of Theoretical Physics}\ }\textbf
		{\bibinfo {volume} {30}},\ \bibinfo {pages} {275} (\bibinfo {year}
		{1963})}\BibitemShut {NoStop}%
	\bibitem [{\citenamefont {Ole{\'{s}}}(1983)}]{Oles1983}%
	\BibitemOpen
	\bibfield  {author} {\bibinfo {author} {\bibfnamefont {A.~M.}\ \bibnamefont
			{Ole{\'{s}}}},\ }\href {\doibase 10.1103/PhysRevB.28.327} {\bibfield
		{journal} {\bibinfo  {journal} {Physical Review B}\ }\textbf {\bibinfo
			{volume} {28}},\ \bibinfo {pages} {327} (\bibinfo {year} {1983})}\BibitemShut
	{NoStop}%
	\bibitem [{\citenamefont {Georgescu}\ and\ \citenamefont
		{Ismail-Beigi}(2015)}]{Georgescu2015}%
	\BibitemOpen
	\bibfield  {author} {\bibinfo {author} {\bibfnamefont {A.~B.}\ \bibnamefont
			{Georgescu}}\ and\ \bibinfo {author} {\bibfnamefont {S.}~\bibnamefont
			{Ismail-Beigi}},\ }\href
	{https://link.aps.org/doi/10.1103/PhysRevB.92.235117} {\bibfield  {journal}
		{\bibinfo  {journal} {Physical Review B}\ }\textbf {\bibinfo {volume} {92}},\
		\bibinfo {pages} {235117} (\bibinfo {year} {2015})}\BibitemShut {NoStop}%
	\bibitem [{\citenamefont {Castellani}\ \emph {et~al.}(1978)\citenamefont
		{Castellani}, \citenamefont {Natoli},\ and\ \citenamefont
		{Ranninger}}]{Castellani1978}%
	\BibitemOpen
	\bibfield  {author} {\bibinfo {author} {\bibfnamefont {C.}~\bibnamefont
			{Castellani}}, \bibinfo {author} {\bibfnamefont {C.~R.}\ \bibnamefont
			{Natoli}}, \ and\ \bibinfo {author} {\bibfnamefont {J.}~\bibnamefont
			{Ranninger}},\ }\href
	{https://journals-aps-org.myaccess.library.utoronto.ca/prb/pdf/10.1103/PhysRevB.18.4945
		https://link.aps.org/doi/10.1103/PhysRevB.18.4945} {\bibfield  {journal}
		{\bibinfo  {journal} {Physical Review B}\ }\textbf {\bibinfo {volume} {18}},\
		\bibinfo {pages} {4945} (\bibinfo {year} {1978})}\BibitemShut {NoStop}%
	\bibitem [{\citenamefont {Johnston}\ \emph {et~al.}(2014)\citenamefont
		{Johnston}, \citenamefont {Mukherjee}, \citenamefont {Elfimov}, \citenamefont
		{Berciu},\ and\ \citenamefont {Sawatzky}}]{Johnston2014}%
	\BibitemOpen
	\bibfield  {author} {\bibinfo {author} {\bibfnamefont {S.}~\bibnamefont
			{Johnston}}, \bibinfo {author} {\bibfnamefont {A.}~\bibnamefont {Mukherjee}},
		\bibinfo {author} {\bibfnamefont {I.}~\bibnamefont {Elfimov}}, \bibinfo
		{author} {\bibfnamefont {M.}~\bibnamefont {Berciu}}, \ and\ \bibinfo {author}
		{\bibfnamefont {G.~A.}\ \bibnamefont {Sawatzky}},\ }\href {\doibase
		10.1103/PhysRevLett.112.106404} {\bibfield  {journal} {\bibinfo  {journal}
			{Physical Review Letters}\ }\textbf {\bibinfo {volume} {112}},\ \bibinfo
		{pages} {106404} (\bibinfo {year} {2014})}\BibitemShut {NoStop}%
	\bibitem [{\citenamefont {Subedi}\ \emph {et~al.}(2015)\citenamefont {Subedi},
		\citenamefont {Peil},\ and\ \citenamefont {Georges}}]{Subedi2015}%
	\BibitemOpen
	\bibfield  {author} {\bibinfo {author} {\bibfnamefont {A.}~\bibnamefont
			{Subedi}}, \bibinfo {author} {\bibfnamefont {O.~E.}\ \bibnamefont {Peil}}, \
		and\ \bibinfo {author} {\bibfnamefont {A.}~\bibnamefont {Georges}},\ }\href
	{\doibase 10.1103/PhysRevB.91.075128} {\bibfield  {journal} {\bibinfo
			{journal} {Physical Review B}\ }\textbf {\bibinfo {volume} {91}},\ \bibinfo
		{pages} {075128} (\bibinfo {year} {2015})}\BibitemShut {NoStop}%
	\bibitem [{\citenamefont {Hellmann}(2015)}]{Hellmann2015}%
	\BibitemOpen
	\bibfield  {author} {\bibinfo {author} {\bibfnamefont {H.}~\bibnamefont
			{Hellmann}},\ }in\ \href {\doibase 10.1007/978-3-662-45967-6_2} {\emph
		{\bibinfo {booktitle} {Hans Hellmann: Einf{\"{u}}hrung in die
				Quantenchemie}}}\ (\bibinfo  {publisher} {Springer Berlin Heidelberg},\
	\bibinfo {address} {Berlin, Heidelberg},\ \bibinfo {year} {2015})\ pp.\
	\bibinfo {pages} {19--376}\BibitemShut {NoStop}%
	\bibitem [{\citenamefont {Feynman}(1939)}]{Feynman1939}%
	\BibitemOpen
	\bibfield  {author} {\bibinfo {author} {\bibfnamefont {R.~P.}\ \bibnamefont
			{Feynman}},\ }\href {\doibase 10.1103/PhysRev.56.340} {\bibfield  {journal}
		{\bibinfo  {journal} {Physical Review}\ }\textbf {\bibinfo {volume} {56}},\
		\bibinfo {pages} {340} (\bibinfo {year} {1939})}\BibitemShut {NoStop}%
	\bibitem [{\citenamefont {Griffiths}(2005)}]{Griffiths2005}%
	\BibitemOpen
	\bibfield  {author} {\bibinfo {author} {\bibfnamefont {D.~J.}\ \bibnamefont
			{Griffiths}},\ }\href@noop {} {\emph {\bibinfo {title} {{Introduction to
					Quantum Mechanics}}}},\ \bibinfo {edition} {2nd}\ ed.\ (\bibinfo  {publisher}
	{Pearson Prentice Hall},\ \bibinfo {address} {Toronto},\ \bibinfo {year}
	{2005})\ pp.\ \bibinfo {pages} {287--288}\BibitemShut {NoStop}%
	\bibitem [{\citenamefont {Jensen}(2007)}]{Jensen2007}%
	\BibitemOpen
	\bibfield  {author} {\bibinfo {author} {\bibfnamefont {F.}~\bibnamefont
			{Jensen}},\ }\href@noop {} {\emph {\bibinfo {title} {{Introduction to
					Computational Chemistry}}}},\ \bibinfo {edition} {2nd}\ ed.\ (\bibinfo
	{publisher} {John Wiley {\&} Sons Inc.},\ \bibinfo {address} {Mississauga},\
	\bibinfo {year} {2007})\ pp.\ \bibinfo {pages} {321--323}\BibitemShut
	{NoStop}%
	\bibitem [{Note2()}]{Note2}%
	\BibitemOpen
	\bibinfo {note} {\protect \textit {i.e.} a cubic equation lacking a quadratic
		term.}\BibitemShut {Stop}%
	\bibitem [{\citenamefont {Mizokawa}\ and\ \citenamefont
		{Fujimori}(1996)}]{Mizokawa1996}%
	\BibitemOpen
	\bibfield  {author} {\bibinfo {author} {\bibfnamefont {T.}~\bibnamefont
			{Mizokawa}}\ and\ \bibinfo {author} {\bibfnamefont {A.}~\bibnamefont
			{Fujimori}},\ }\href {\doibase 10.1103/PhysRevB.54.5368} {\bibfield
		{journal} {\bibinfo  {journal} {Physical Review B}\ }\textbf {\bibinfo
			{volume} {54}},\ \bibinfo {pages} {5368} (\bibinfo {year}
		{1996})}\BibitemShut {NoStop}%
	\bibitem [{\citenamefont {Peters}(2009)}]{Peters2009a}%
	\BibitemOpen
	\bibfield  {author} {\bibinfo {author} {\bibfnamefont {R.}~\bibnamefont
			{Peters}},\ }\emph {\bibinfo {title} {{Magnetic Phases in the Hubbard
				Model}}},\ \href
	{https://ediss.uni-goettingen.de/bitstream/handle/11858/00-1735-0000-000D-F17A-4/peters.pdf?sequence=1}
	{Ph.D. thesis},\ \bibinfo  {school} {der Georg-August-Universit\"at
		G\"ottingen} (\bibinfo {year} {2009})\BibitemShut {NoStop}%
	\bibitem [{\citenamefont {Banerjee}\ \emph {et~al.}(2016)\citenamefont
		{Banerjee}, \citenamefont {Suryanarayana},\ and\ \citenamefont
		{Pask}}]{Banerjee2016}%
	\BibitemOpen
	\bibfield  {author} {\bibinfo {author} {\bibfnamefont {A.~S.}\ \bibnamefont
			{Banerjee}}, \bibinfo {author} {\bibfnamefont {P.}~\bibnamefont
			{Suryanarayana}}, \ and\ \bibinfo {author} {\bibfnamefont {J.~E.}\
			\bibnamefont {Pask}},\ }\href {\doibase 10.1016/j.cplett.2016.01.033}
	{\bibfield  {journal} {\bibinfo  {journal} {Chemical Physics Letters}\
		}\textbf {\bibinfo {volume} {647}},\ \bibinfo {pages} {31} (\bibinfo {year}
		{2016})}\BibitemShut {NoStop}%
	\bibitem [{\citenamefont {Pulay}(1980)}]{Pulay1980}%
	\BibitemOpen
	\bibfield  {author} {\bibinfo {author} {\bibfnamefont {P.}~\bibnamefont
			{Pulay}},\ }\href {\doibase 10.1016/0009-2614(80)80396-4} {\bibfield
		{journal} {\bibinfo  {journal} {Chemical Physics Letters}\ }\textbf {\bibinfo
			{volume} {73}},\ \bibinfo {pages} {393} (\bibinfo {year} {1980})}\BibitemShut
	{NoStop}%
	\bibitem [{\citenamefont {Mazin}\ \emph {et~al.}(2007)\citenamefont {Mazin},
		\citenamefont {Khomskii}, \citenamefont {Lengsdorf}, \citenamefont {Alonso},
		\citenamefont {Marshall}, \citenamefont {Ibberson}, \citenamefont
		{Podlesnyak}, \citenamefont {Mart{\'{i}}nez-Lope},\ and\ \citenamefont
		{Abd-Elmeguid}}]{Mazin2007}%
	\BibitemOpen
	\bibfield  {author} {\bibinfo {author} {\bibfnamefont {I.~I.}\ \bibnamefont
			{Mazin}}, \bibinfo {author} {\bibfnamefont {D.~I.}\ \bibnamefont {Khomskii}},
		\bibinfo {author} {\bibfnamefont {R.}~\bibnamefont {Lengsdorf}}, \bibinfo
		{author} {\bibfnamefont {J.~A.}\ \bibnamefont {Alonso}}, \bibinfo {author}
		{\bibfnamefont {W.~G.}\ \bibnamefont {Marshall}}, \bibinfo {author}
		{\bibfnamefont {R.~M.}\ \bibnamefont {Ibberson}}, \bibinfo {author}
		{\bibfnamefont {A.}~\bibnamefont {Podlesnyak}}, \bibinfo {author}
		{\bibfnamefont {M.~J.}\ \bibnamefont {Mart{\'{i}}nez-Lope}}, \ and\ \bibinfo
		{author} {\bibfnamefont {M.~M.}\ \bibnamefont {Abd-Elmeguid}},\ }\href
	{\doibase 10.1103/PhysRevLett.98.176406} {\bibfield  {journal} {\bibinfo
			{journal} {Physical Review Letters}\ }\textbf {\bibinfo {volume} {98}},\
		\bibinfo {pages} {176406} (\bibinfo {year} {2007})}\BibitemShut {NoStop}%
	\bibitem [{\citenamefont {Peters}\ and\ \citenamefont
		{Pruschke}(2009)}]{Peters2009b}%
	\BibitemOpen
	\bibfield  {author} {\bibinfo {author} {\bibfnamefont {R.}~\bibnamefont
			{Peters}}\ and\ \bibinfo {author} {\bibfnamefont {T.}~\bibnamefont
			{Pruschke}},\ }\href {\doibase 10.1103/PhysRevB.79.045108} {\bibfield
		{journal} {\bibinfo  {journal} {Physical Review B}\ }\textbf {\bibinfo
			{volume} {79}},\ \bibinfo {pages} {045108} (\bibinfo {year} {2009})}\BibitemShut
	{NoStop}%
	\bibitem [{\citenamefont {Eckstein}\ \emph {et~al.}(2005)\citenamefont
		{Eckstein}, \citenamefont {Kollar}, \citenamefont {Byczuk},\ and\
		\citenamefont {Vollhardt}}]{Eckstein2005}%
	\BibitemOpen
	\bibfield  {author} {\bibinfo {author} {\bibfnamefont {M.}~\bibnamefont
			{Eckstein}}, \bibinfo {author} {\bibfnamefont {M.}~\bibnamefont {Kollar}},
		\bibinfo {author} {\bibfnamefont {K.}~\bibnamefont {Byczuk}}, \ and\ \bibinfo
		{author} {\bibfnamefont {D.}~\bibnamefont {Vollhardt}},\ }\href {\doibase
		10.1103/PhysRevB.71.235119} {\bibfield  {journal} {\bibinfo  {journal}
			{Physical Review B}\ }\textbf {\bibinfo {volume} {71}},\ \bibinfo {pages}
		{235119} (\bibinfo {year} {2005})}\BibitemShut {NoStop}%
	\bibitem [{\citenamefont {Mizokawa}\ \emph {et~al.}(1995)\citenamefont
		{Mizokawa}, \citenamefont {Fujimori}, \citenamefont {Arima}, \citenamefont
		{Tokura}, \citenamefont {Mori},\ and\ \citenamefont
		{Akimitsu}}]{Mizokawa1995}%
	\BibitemOpen
	\bibfield  {author} {\bibinfo {author} {\bibfnamefont {T.}~\bibnamefont
			{Mizokawa}}, \bibinfo {author} {\bibfnamefont {A.}~\bibnamefont {Fujimori}},
		\bibinfo {author} {\bibfnamefont {T.}~\bibnamefont {Arima}}, \bibinfo
		{author} {\bibfnamefont {Y.}~\bibnamefont {Tokura}}, \bibinfo {author}
		{\bibfnamefont {N.}~\bibnamefont {Mori}}, \ and\ \bibinfo {author}
		{\bibfnamefont {J.}~\bibnamefont {Akimitsu}},\ }\href {\doibase
		10.1103/PhysRevB.52.13865} {\bibfield  {journal} {\bibinfo  {journal}
			{Physical Review B}\ }\textbf {\bibinfo {volume} {52}},\ \bibinfo {pages}
		{13865} (\bibinfo {year} {1995})}\BibitemShut {NoStop}%
\end{thebibliography}

\end{document}